\documentclass[fleqn,usenatbib]{mnras}

\usepackage{newtxtext,newtxmath}
\usepackage[T1]{fontenc}

\DeclareRobustCommand{\VAN}[3]{#2}
\let\VANthebibliography\thebibliography
\def\thebibliography{\DeclareRobustCommand{\VAN}[3]{##3}\VANthebibliography}

\usepackage{graphicx}	% Including figure files
\usepackage{amsmath}	% Advanced maths commands
\usepackage{wrapfig}
\usepackage[utf8]{inputenc}
\usepackage{makecell}
\usepackage[utf8]{inputenc}
\usepackage[french]{babel}
\usepackage{lscape}
\usepackage[flushleft]{threeparttable}
\usepackage{pdflscape}
\usepackage{gensymb}
\usepackage{longtable}
\usepackage[export]{adjustbox}

\title[SITELLE view of galaxies Arp 82]{The interacting pair of galaxies Arp 82: Integral field spectroscopy and numerical simulations}

\author[P. Karera et al.]
{Prime Karera,$^{1,2}$\thanks{E-mail: prime.karera.1@ulaval.ca (KTS)}
Laurent Drissen,$^{1,2}$
Hugo Martel,$^{1,2}$
Jorge Iglesias-P\'aramo,$^{3}$
Jose M. Vilchez,$^{3}$
\newauthor Pierre-Alain Duc,$^{4}$
and Henri Plana$^{5}$
\\
\\
$^{1}$Département de Physique, de génie physique et d'optique, Université Laval, Québec (QC), G1V 0A6, Canada\\
$^{2}$Centre de recherche en astrophysique du Québec, Québec (QC), Canada\\
$^{3}$Instituto de Astrofísica de Andalucía (CSIC), Glorieta de la Astronomía s/n, Aptdo. 3004, 18080 Granada, Spain\\
$^{4}$Observatoire Astronomique de Strasbourg (ObAS)
11, rue de l'Universit\'e, F-67000 Strasbourg, France\\
$^{5}$Laboratório de Astrofísica Teórica e Observacional, Universidade Estadual de Santa Cruz, 45650-000 Ilhéus, BA, Brasil\\
}

\date{Accepted 2022 May 23. Received 2022 May 22; in original form 2022 February 14}

\pubyear{2020}

\begin{document}
\label{firstpage}
\pagerange{\pageref{firstpage}--\pageref{lastpage}}
\maketitle

\begin{abstract}
Spectral data cubes of the interacting pair of galaxies NGC 2535 and NGC 2536 (the Arp 82 system) targeting bright emission lines in the visible band, obtained with the imaging Fourier transform spectrometer (iFTS) SITELLE attached to the Canada-France-Hawaii Telescope (CFHT), are presented. Analysis of H$\upalpha$ velocity maps reveals a bar in $\rm NGC\,2536$. In $\rm NGC\,2535$, we find strong non-circular motions outside the ocular ring, in the elliptical arc and tidal tails of $\rm NGC\,2535$ and a misalignment between the kinematic and photometric position angles. We detect 155 HII region complexes in the interacting pair of galaxies and determine oxygen abundances for 66 of them using different calibrators. We find, regardless of the indicator used, that the oxygen abundance distribution in $\rm NGC\,2536$ is shallow whereas, in $\rm NGC\,2535$, it is best fitted by two slopes, the break occurring beyond the ocular ring. The inner slope is comparable to the one observed in isolated normal star-forming galaxies but the outer slope is shallow. We present a numerical simulation of the interaction that reproduces the observed tidal features, kinematics, and metallicity distribution, to investigate the effect of the interaction on the galaxies. The model indicates that the galaxies have undergone a close encounter, strongly prograde for the primary, and are half way in their course to a second close encounter.
\end{abstract}


%%%%%%%%%%%%%%%%%%%%%%%%%%%%%%%%%%%%%%%%%%%%%%%%%%

%%%%%%%%%%%%%%%%% BODY OF PAPER %%%%%%%%%%%%%%%%%%

\section{Introduction}
Galaxy interactions are fundamental in the evolution of the Universe. Early numerical simulations \citep[][]{Toomre&Toomre1972,Toomre1977} demonstrated that gravitational forces between merging or interacting galaxies were responsible for the distorted morphologies observed in these peculiar systems. Later, simulations taking into account the hydrodynamics of the gaseous component \citep[e.g.,][]{Mihos1994,DiMatteo2007,Blumenthal2018}, suggested that tidal forces may funnel gas to the central regions and fuel massive star formation (SF) and/or active galactic nuclei (AGN). These theoretical predictions are supported by observational studies \citep[][and references therein]{Ellison2013}. Specifically, ultraluminous infrared galaxies (ULIRGs) at low redshifts, i.e the strongest starbursts in the local Universe, are found preferentially in merging and interacting galaxies \citep[][]{Sanders&Mirabel1996,Knapen2015}. The inflow of metal-poor gas towards the centre of the galaxy dilutes the nuclear metallicity and flattens the initial metallicity gradient as has been observed in some nearby galaxy pairs \citep[][]{Kewley2006,Kewley2010} and reproduced by numerical simulations \citep[][]{Rupke2010}.

However, induced SF is not always localized in interacting galaxies. It can be spatially extended \citep[][]{Renaud2015} and, under favorable conditions, gravitationally bound tidal dwarf galaxies (TDGs) can form up to the tip of tidal tails \citep[][]{Duc2012}. The magnitude of induced changes strongly depend on the interaction stage, orbital parameters and intrinsic properties of the colliding galaxies. For instance, distant passages (usually at an early stage of interaction) trigger an increase in the SF activity over large volumes, whereas, closer, penetrating encounters (usually at later stages before coalescence) increase nuclear activity. Moreover, \citet{Barrera-Ballesteros2015} report central oxygen abundances in a sample of interacting galaxies similar to those in a control sample, suggesting that other processes such as stellar feedback contribute to lower central metallicity depression in interacting galaxies. Therefore, to understand galaxy-galaxy interactions, a combination of 2D spectroscopic and imaging data, which yields simultaneous information on morphology, SF, metallicity, etc. is needed. Comparison of observations to numerical models of the interactions can provide information about orbital parameters, timescales and SF triggering mechanisms \citep[][]{Renaud2015}. At a minimum, a numerical model should reproduce the morphology and kinematics of the system \citep[][]{Privon2013}. Insights on metallicity distribution would further constrain the numerical model. To compare the velocity fields of simulations and observations, we need to have relatively high velocity resolution and spatial resolution for the systems being modelled. There are only a few systems where these kinematic data with high resolutions and spatially covering both the parents disk galaxies, the tidal features and possible TDGs are available \citep[][]{Bournaud2004}. 

In this article, we are aiming to investigate the impact of interactions on the properties of the ionized gas in the pair of galaxies Arp 82, an M51-like double system that consists of $\rm NGC\,2535$, a galaxy with grand-design spiral arms and a companion, $\rm NGC\,2536$, at the end of one of the tidal tails. $\rm NGC\,2535$ has an eye-shaped structure connecting the two tidal arms reminiscent of the ocular galaxy IC\,2163. 2D velocity maps were obtained by \citet{Amram1989} in H$\upalpha$ and by \citet{Kaufman1997} in HI line. 1D velocity and metallicity measurements were also obtained by \citet{Zasov2019} using two long slits, one connecting the centres of the galaxies and the other running along the bridge and crossing $\rm NGC\,2536$. Using broadband UV and optical colors, combined to population synthesis models, \citet{Hancock2007} find that the stellar population is essentially of young to intermediate ages. Numerical models that successfully reproduce the morphology of the system \citep[][]{Howard1993, Kaufman1997, Hancock2007, Holincheck2016} agree on the fact that the interaction is strongly prograde for the primary galaxy but differ, for example, in mass ratios and time elapsed since closest approach. 

In this work, we study the ionized gas properties of Arp 82 using SITELLE data cubes and present a numerical SPH model of the interacting galaxies. We adopt a distance of 59.2 Mpc from \citet{Zaragoza2018}; at this distance, $1^{\prime\prime}$ $\sim 285$ pc. 
With its large field of view ($11^\prime\times11^\prime$) and flexible spectral resolution up to $\rm R=10000$ in filter-selected regions of the visible range, the CFHT imaging Fourier transform spectrometer (iFTS) SITELLE \citep[][]{Drissen2014,Drissen2019} is very well-suited to tackle this problem; its spatial resolution is seeing-limited and sampled at 0.32$^{\prime\prime}$ per pixel. At the distance of Arp 82, the instrument allows a spatial resolution of 215 pc ($0.75^{\prime\prime}$ seeing) up to $\sim188\,\rm kpc$ from the central galaxies.
We derive rotation curves, surface brightness, and oxygen abundance profiles for both galaxies and discuss the evolution of the SF activity and metallicity during the interaction.  

This paper is organised as follows. In Section~\ref{sec:section2} we describe our observations, data reduction and calibration. In Section~\ref{sec:section3} we describe our data analysis. We present our numerical model in Section~\ref{sec:section4}. Finally, we summarize our results in Section~\ref{sec:section5}.

\section{OBSERVATIONS, DATA REDUCTION AND CALIBRATION}
\label{sec:section2}
\begin{figure}
    \includegraphics[width=\columnwidth]{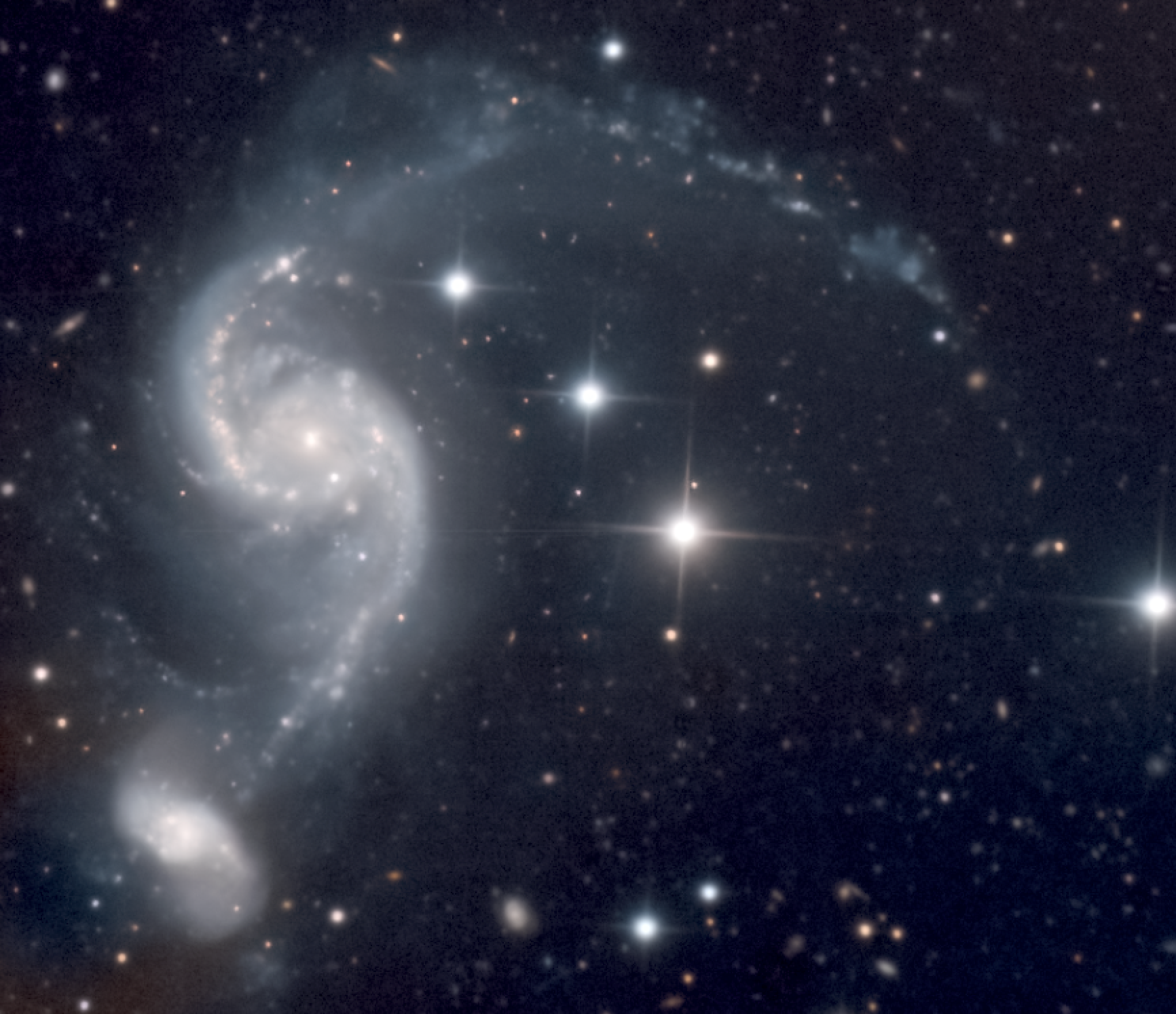}
    \caption{Composite colored image of Arp 82 obtained with SITELLE (blue: SN2 deep image, red: SN3 deep image, green: mean of SN2 and SN3 deep images). North is up and East is left. The image spans $4.9' \times 4.6'$.}
    \label{fig:deepimages_figure}
\end{figure}

Arp 82 was observed with SITELLE  in November 2016 in three spectral bands allowing measurements of the strong emission lines: SN1 filter for [OII]$\uplambda\uplambda$3727,3729, SN2 filter for H$\upbeta$ and [OIII]$\uplambda\uplambda$4959,5007, and SN3 filter for H$\upalpha$, [NII]$\uplambda\uplambda$6548,6583 and [SII]$\uplambda\uplambda$6716,6731). Details and technical specifications of the observations are listed in Table~\ref{tab:table1}.  Fig.~\ref{fig:deepimages_figure} shows a combination of the deep images for the SN2 and SN3 cubes; these images are produced by summing all interferograms obtained by the instrument before performing the Fourier transform. Features such as the bright eye-shaped structure in $\rm NGC\,2535$ and the extended tidal arms are easily seen. 

\begin{table}
    \caption{Observation Log.}\label{tab:table1}
    \centering
    \small
    \begin{tabular}{ccccc}
    \hline
    \small\makecell{Filter} & 
    \small \makecell{Band \\
    \small [\AA]}
    & 
    \small \makecell{Spectral \\ resolution} & 
    \small \makecell{Exp. time \\
    per step \\
    \small [sec]}
    & 
    \small \makecell{\# \\
    of steps} \\
    \hline
    \small SN1 & \small 3650 - 3850 & \small 960 & \small 39 & \small 172 \\
    \hline
    \small SN2 & \small 4800 - 5800 & \small 970 & \small 40 & \small 225 \\
    \hline
    \small SN3 & \small 6510 - 6850 & \small 1900 & \small 25 & \small 337 \\
    \hline
    \end{tabular}
    \end{table}

\begin{figure}
    \includegraphics[width=\columnwidth]{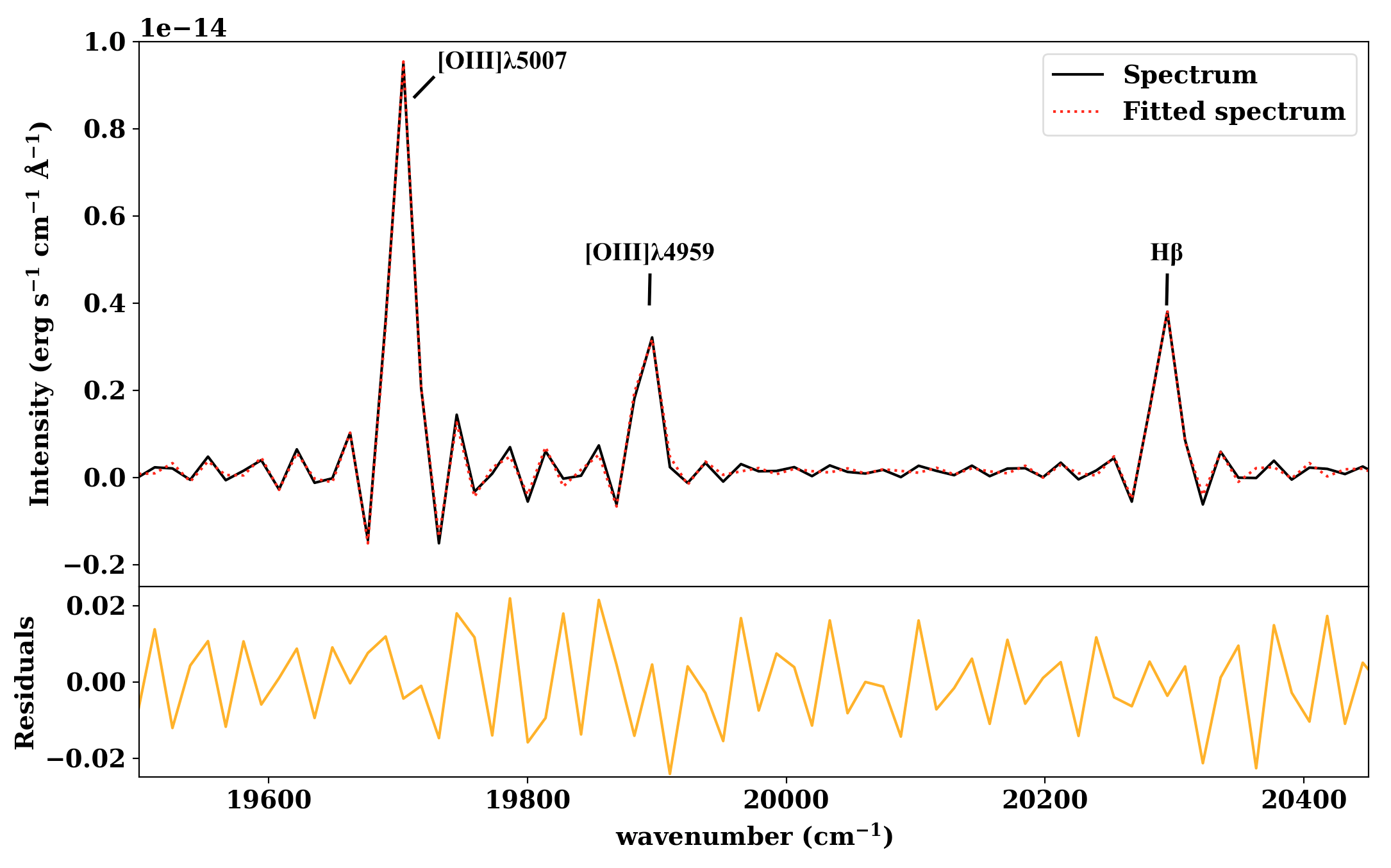}
    \caption{Example of a fit of the spectrum of a bright star-forming region in NGC\,2535. The good quality of the fit of emission lines by sinc-shaped profiles is evidenced by the very small residuals.}
    \label{fig:fitspectrumsn2_figure}
\end{figure}

\begin{figure*}
    \includegraphics[scale=0.8]{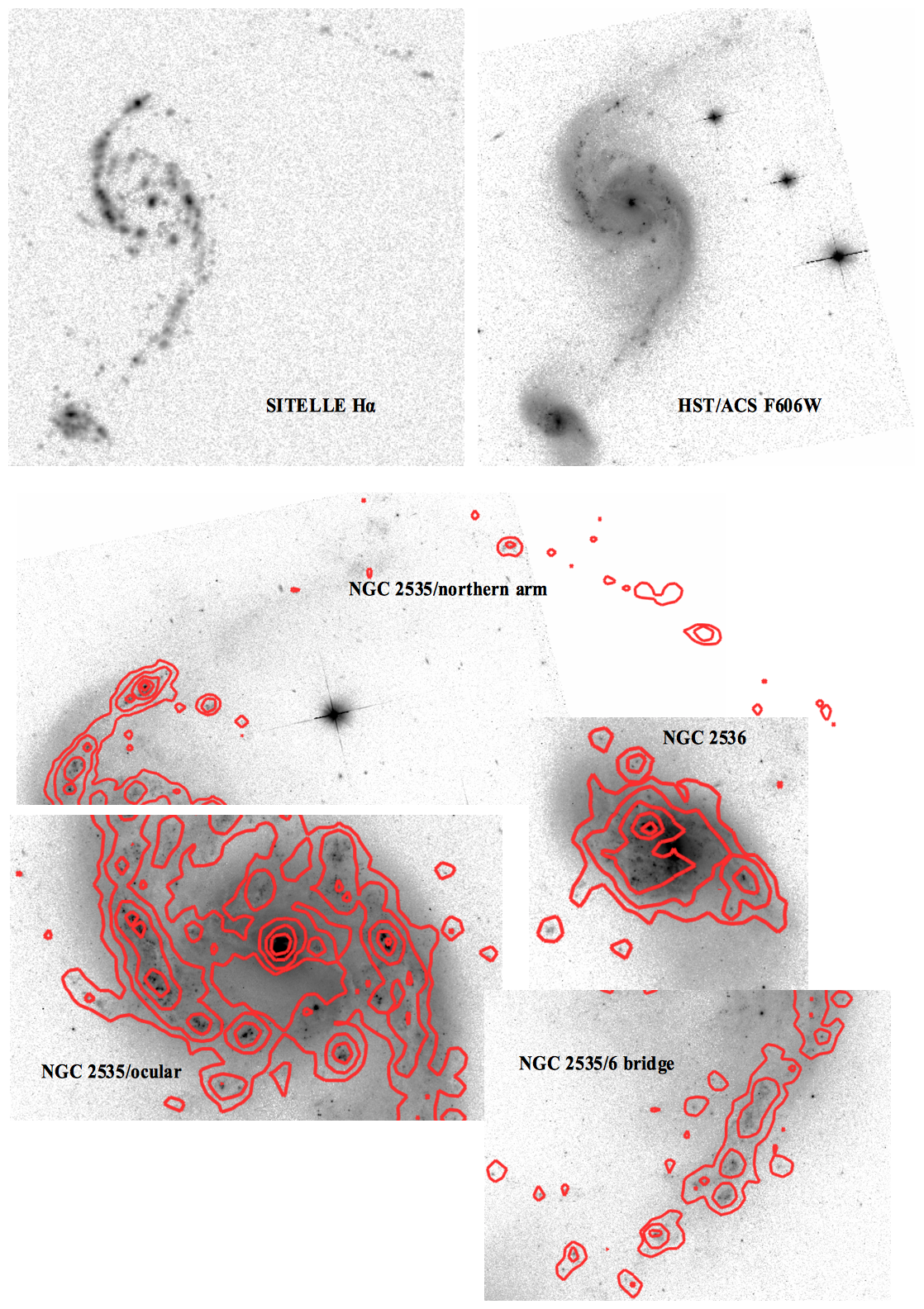}
    \caption{Expanded view of H$\upalpha$ flux map at SITELLE full resolution (top left) and archival HST ACS/F606W image (top right). Zoom in view of the HST image with SITELLE H$\upalpha$ contours overplotted (mosaic in lower panels) : there is a correlation of H$\upalpha$ emission to the presence of ionizing young stellar sources. H$\upalpha$ contour levels are $[5,16,63,239,504]\times10^{-18}\,\rm erg\,s^{-1}\,cm^{-2}$.}
    \label{fig:Hafullresolution_figure}
\end{figure*}

\begin{figure*}
    \includegraphics[width=\textwidth]{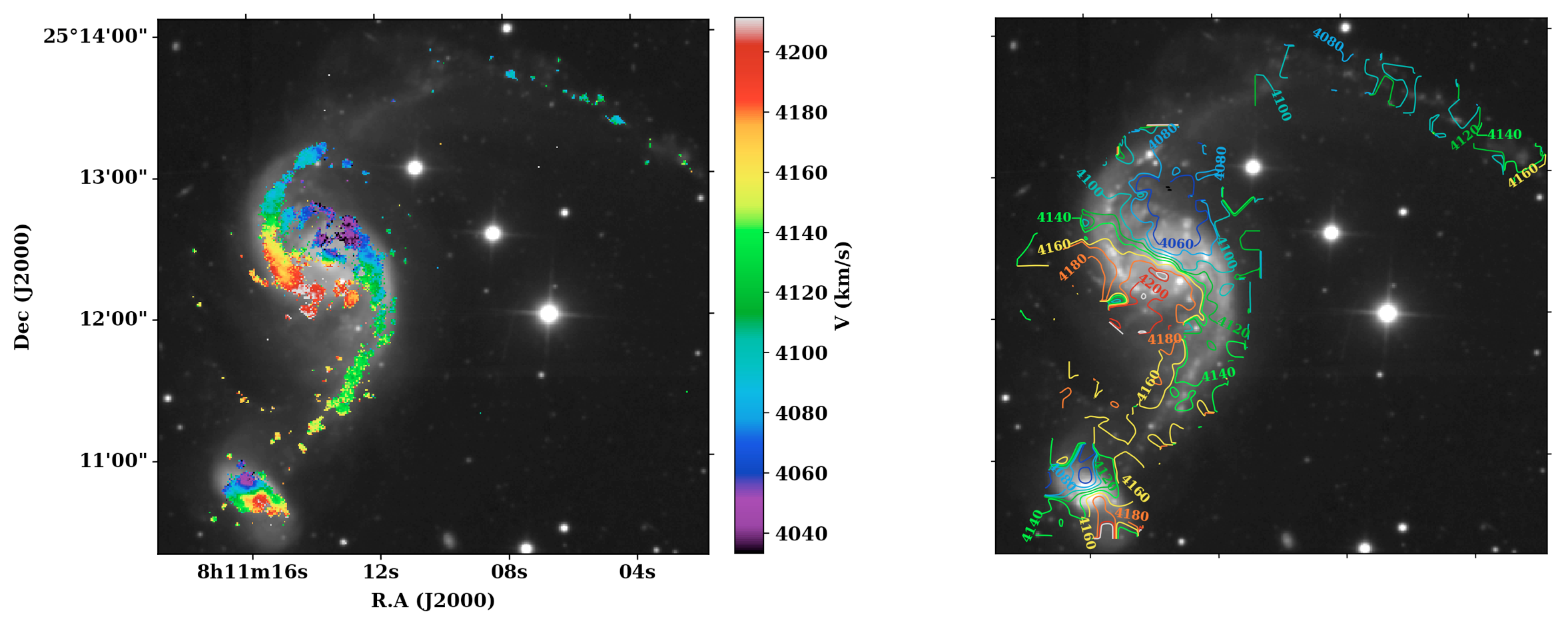}
    \caption{H$\upalpha$ velocity map (left) and corresponding isovelocity contours (right) superimposed on the SN3 deep image. The corresponding map of velocity uncertainty is shown in Fig.~\ref{fig:emissionlines_figure}.Contours were generated from the velocity maps smoothed with a gaussian 2D kernel of 2.5 pixels.}
    \label{fig:velocitymap_figure}
\end{figure*}

Data reduction was performed using ORBS and ORCS \citep[][]{Martin2015}, SITELLE's dedicated data reduction and analysis pipelines. Details of the automated procedure as well as the sources of uncertainties can be found in \citet{Martin2021}. The wavelength calibration is performed using a high resolution laser source but may show distortions (up to 25 km/s across the field of view) due to aberrations and deformations in the optical structure \citep[][]{Martin2016}. To refine the wavelength calibration, we measured the centroid positions of the night sky OH emission lines (in the SN3 filter) in spectra extracted from circular regions sampling the entire FOV, following the procedure described in \citet{Martin2018}. A velocity correction map is obtained by interpolating values on the night sky velocity grid, reducing the uncertainty on the velocity calibration across the entire field to less than 2 km/s. We also applied a barycentric correction of +24.71 km/s. Photometric calibration was performed using images and datacubes of spectrophotometric standard stars.

Given that the spatial resolution is seeing-limited ($\sim0.75^{\prime\prime}$), the spaxels have been binned $2\times2$ during the line extraction process to increase the signal-to-noise ratio (SNR) of faint regions. Emission lines are fitted for each spaxel after subtracting a median sky spectrum taken in a dark circular region (aperture of 13$^{\prime\prime}$) far away from the galaxies. Line fitting for SN1 and SN2 data cubes are done using sinc-shaped profiles, the natural instrument line shape of iFTS, as illustrated in Fig.~\ref{fig:fitspectrumsn2_figure}. The higher spectral resolution and SNR achieved for the SN3 data cube allows us to fit these lines using a sincgauss function (a sinc function convolved with a gaussian; see \citet{Martin2016} for details) and thus map the velocity dispersion of the ionized gas. The software produces maps of fluxes of fitted emission lines, continuum level, velocity, and velocity dispersion (the last three parameters are common to all lines in one data cube) together with maps of their corresponding uncertainties. While the SN1 and SN2 cubes were obtained under photometric conditions, it was not the case for the SN3 cube. We therefore compared our H$\upalpha$+[NII] fluxes with those obtained by \citet{Kennicutt1987}. SN3 flux maps (H$\upalpha$, [NII]$\uplambda\uplambda$6548,6583, [SII]$\uplambda\uplambda$6716,6731) are thereafter corrected to match \citet{Kennicutt1987} data. Flux maps of extracted emission lines and the velocity dispersion map from the SN3 cube are presented in Fig.~\ref{fig:emissionlines_figure}. For illustration, we show in Fig.~\ref{fig:Hafullresolution_figure} (top left) the H$\upalpha$ flux map at full resolution. Several bright HII region complexes can be identified in both galaxies, the faintest regions being located in the tidal tails far from the galaxies centres. This H$\upalpha$ emission results from young stellar sources ionizing their surrounding interstellar medium. The correlation can be seen in the zoom in snapshots of Fig.~\ref{fig:Hafullresolution_figure} where peaks of H$\upalpha$ emission coincides with loci of bright star-forming knots visible in the sharp archival Hubble Space Telescope (HST) Advanced Camera for Surveys (ACS) image (filter F606W). Star forming regions, as revealed in SITELLE and HST images of Fig.~\ref{fig:Hafullresolution_figure}, are not only confined in the centres of the galaxies. Both images reveal that star formation rate (SFR) is greater in the ocular and spiral structures and lower in the tidal tails (as shall be quantified in Section~\ref{sec:section3} from H$\upalpha$ fluxes corrected for extinction), although the northern tail of $\rm NGC\,2535$ is truncated in the HST image. We also note that the H$\upalpha$ flux in the central region of NGC 2536 (including the bar) is relatively weak, the brightest HII region complexes being clearly offset from the core of the galaxy.
The analysis in the following section is based on the spectral properties of the ionized gas in the interacting system.

\section{DATA ANALYSIS}
\label{sec:section3}
\subsection{Kinematics and surface brightness}
\label{sec:section31}

The line-of-sight velocity maps obtained from fitting SN3 emission lines of the ionized gas are presented in Fig.~\ref{fig:velocitymap_figure}. Only data points with ${\rm H\upalpha\ fluxes}>2\times10^{-17}\rm erg\,s^{-1}\,cm^{-2}$, $\rm SNR > 3$ and measured velocity uncertainties < 20 km s$^{-1}$ were used. The isovelocity contours, although very distorted, show an overall "spider" pattern characteristic of differentially rotating inclined disks. For both galaxies, approaching/receding sides have projected velocities lesser/greater than 4100/4140 km\,s$^{-1}$, with systemic velocities $\sim$ 4120 km\,s$^{-1}$. The line of nodes in $\rm NGC\,2535$ lies in the direction joining the centres of the two galaxies; in $\rm NGC\,2536$ it lies in the direction joining the tips of the spiral arms. Assuming that both galaxies have trailing arms, they both rotate counterclockwise. 

The velocity fields of the galaxies are very perturbed. Isovelocity contours tend to run parallel to spiral arms with velocity kinks indicative of peculiar velocities or streaming motions. The velocity kinks are also observed in the HI velocity map of \citet{Kaufman1997} in which an elliptical arc is identified with a position angle (P.A.) of $172\degree$. This feature coincides in our velocity map to the loci of velocity kinks around $\rm NGC\,2535$ delimited to the south-east by the curved spur of HII regions right above $\rm NGC\,2536$ and to the north-west by the string of HII regions intersecting the tidal arm at $\sim$ [8h11m14s,$25\degree13^{\prime}00^{\prime\prime}$]. The ellipse is off-centered as isovelocity contours are more elongated on the receding side. The ocular ring and the smaller spurs that appear attached to the ring at its southern rim and the one right above the ring's northern rim are not seen in the HI map as the central part of $\rm NGC\,2535$ is located in an HI trough filled in with $^{12}\rm CO$ emission \citep[][]{Kaufman1997}. The link between these features and the ongoing interaction between the two galaxies is discussed in Section~\ref{sec:section4}.

\begin{figure*}
    \includegraphics[scale=0.9]{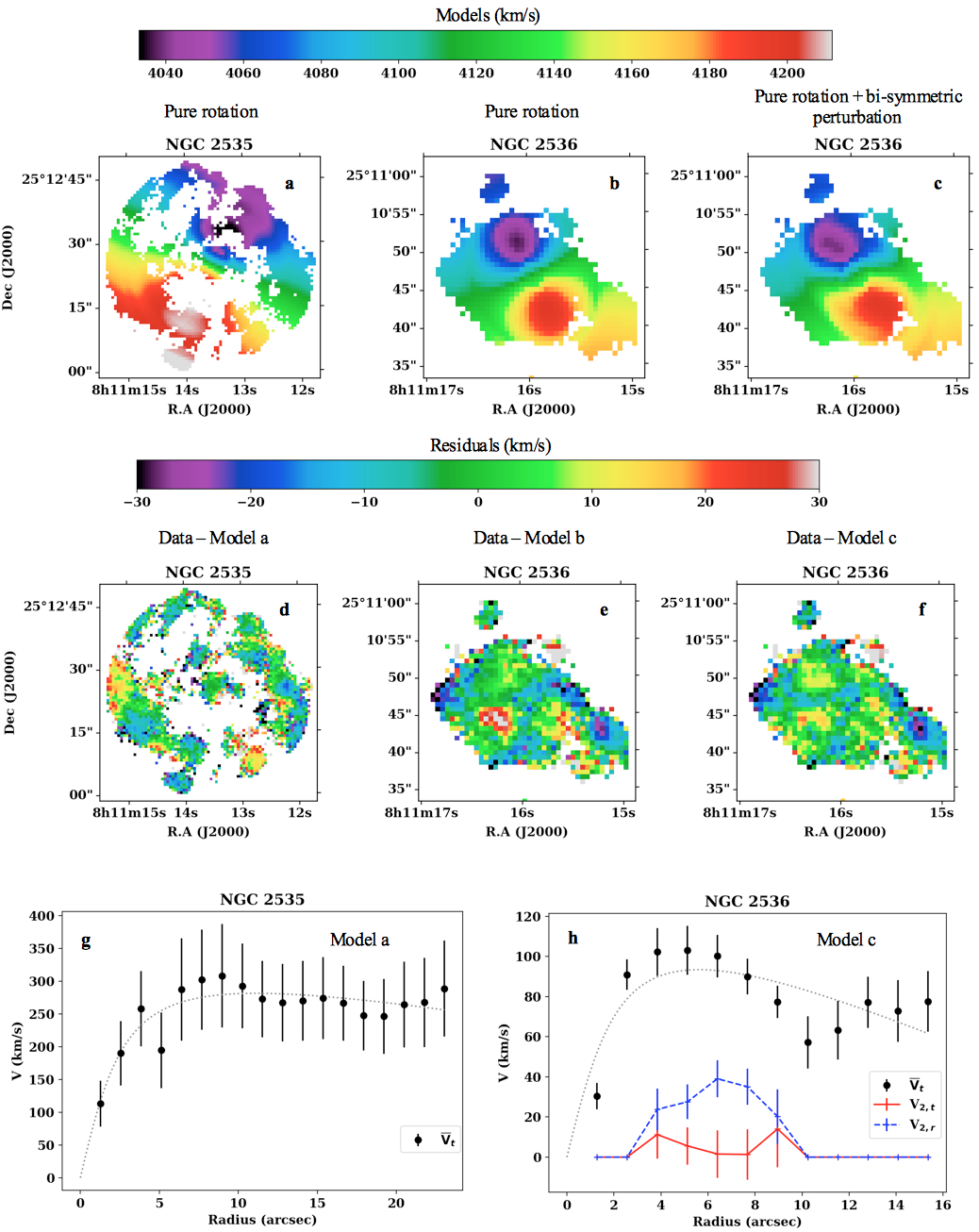}
    \caption{Fitted velocity models for the H$\upalpha$ velocity fields. (a - c): Rotation-only models for $\rm NGC\,2535$ and $\rm NGC\,2536$ and rotation+bi-symmetric flow model for $\rm NGC\,2536$. (d - f): corresponding residuals. (g, h): Rotation curves correspond to the best-fitting models, i.e rotation-only model for $\rm NGC\,2535$ and rotation+bi-symmetric flow for $\rm NGC\,2536$. The dotted lines are arctan functional fits to the rotation curves (see text).}
    \label{fig:Diskfitvelmod_figure}
\end{figure*}

\begin{figure*}
    \centering
    \includegraphics[scale=0.55]{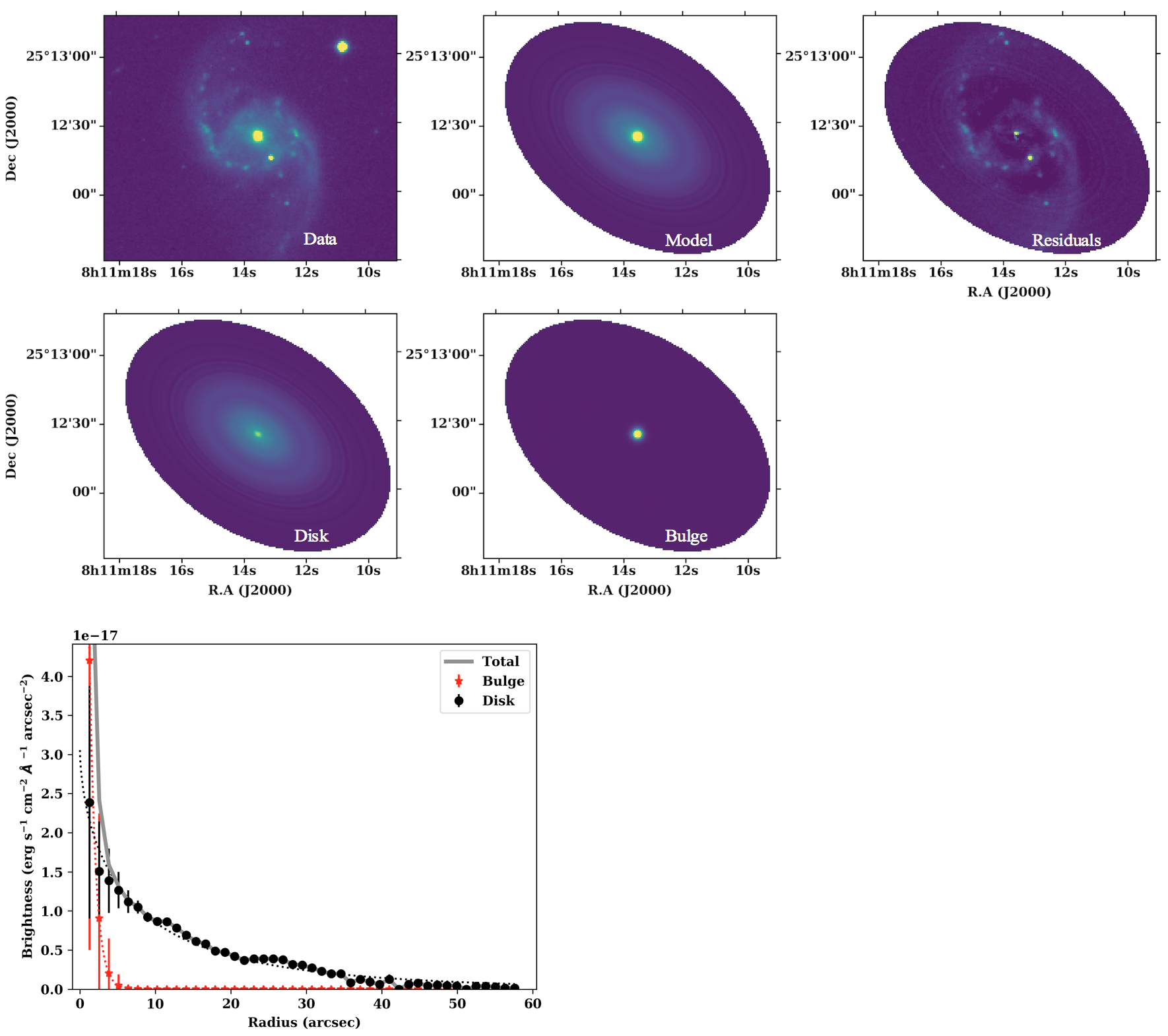}
    \caption{SN3 band continuum model for $\rm NGC\,2535$. Top left: SN3 continuum image of $\rm NGC\,2535$ ; top centre: disk+bulge DiskFit model; top right: (data\text{--}model) residuals; middle left: disk component of the model; middle centre: bulge component of the model; bottom: radial brightness distribution of the model components. Dashed lines are fits to brightness profiles with an exponential function (Eq.~\eqref{eq:bulgemodel})}
    \label{fig:Diskfitphotmod2535_figure}
\end{figure*}

\begin{figure*}
    \centering
    \includegraphics[scale=0.55]{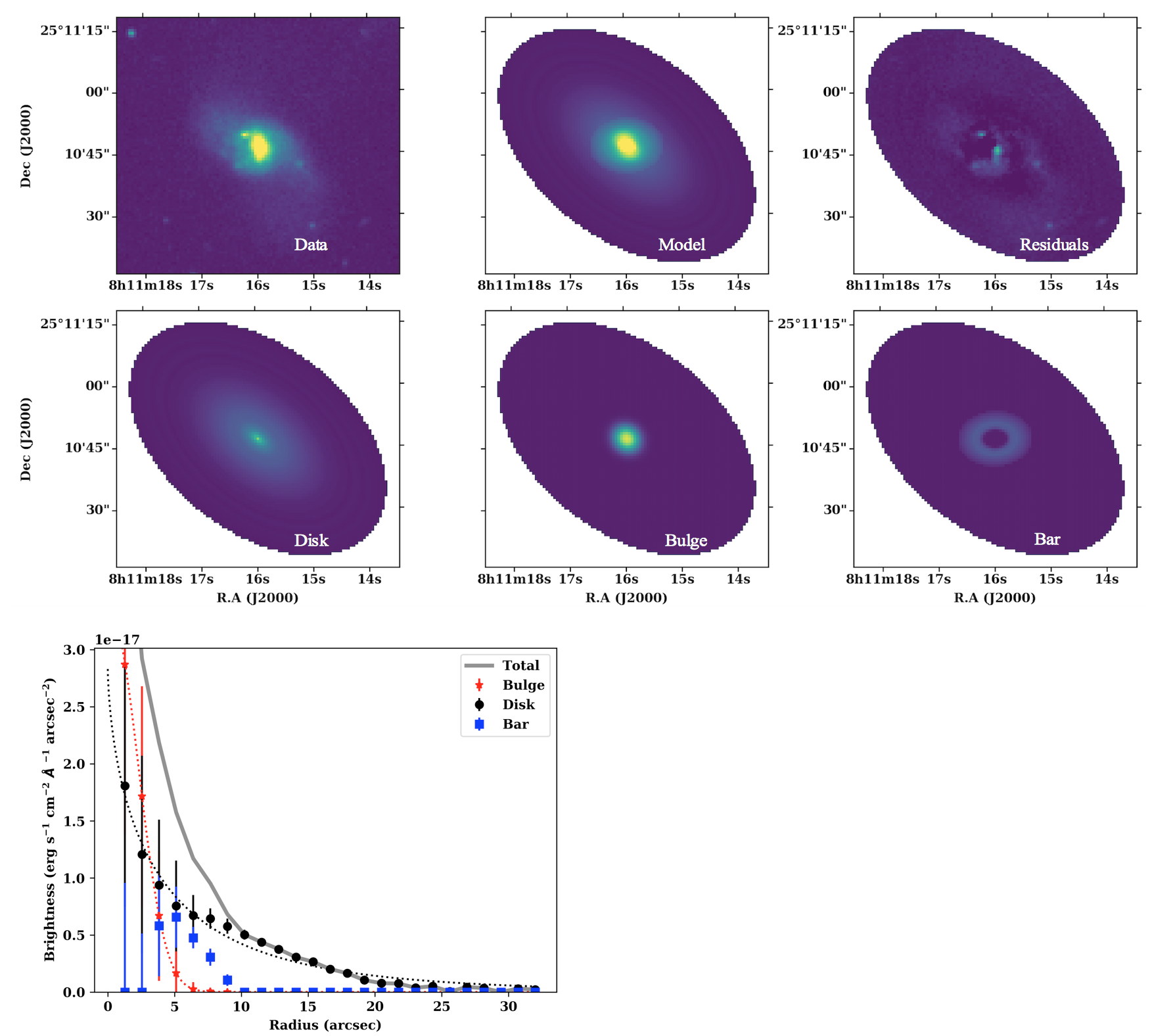}
    \caption{Same as in Fig.~\ref{fig:Diskfitphotmod2535_figure} but for $\rm NGC\,2536$. Top left: SN3 continuum image of $\rm NGC\,2536$ ; top centre: disk+bulge+bar DiskFit model; top right: (data\text{--}model) residuals; middle left: disk component of the model; middle centre: bulge component of the model; middle left: bar component of the model; bottom: radial brightness distribution of the model components. Dashed lines are fits to brightness profiles with an exponential function (Eq.~\eqref{eq:bulgemodel})}
    \label{fig:Diskfitphotmod2536_figure}
\end{figure*}

\begin{figure*}
    \centering
    \includegraphics[scale=0.68]{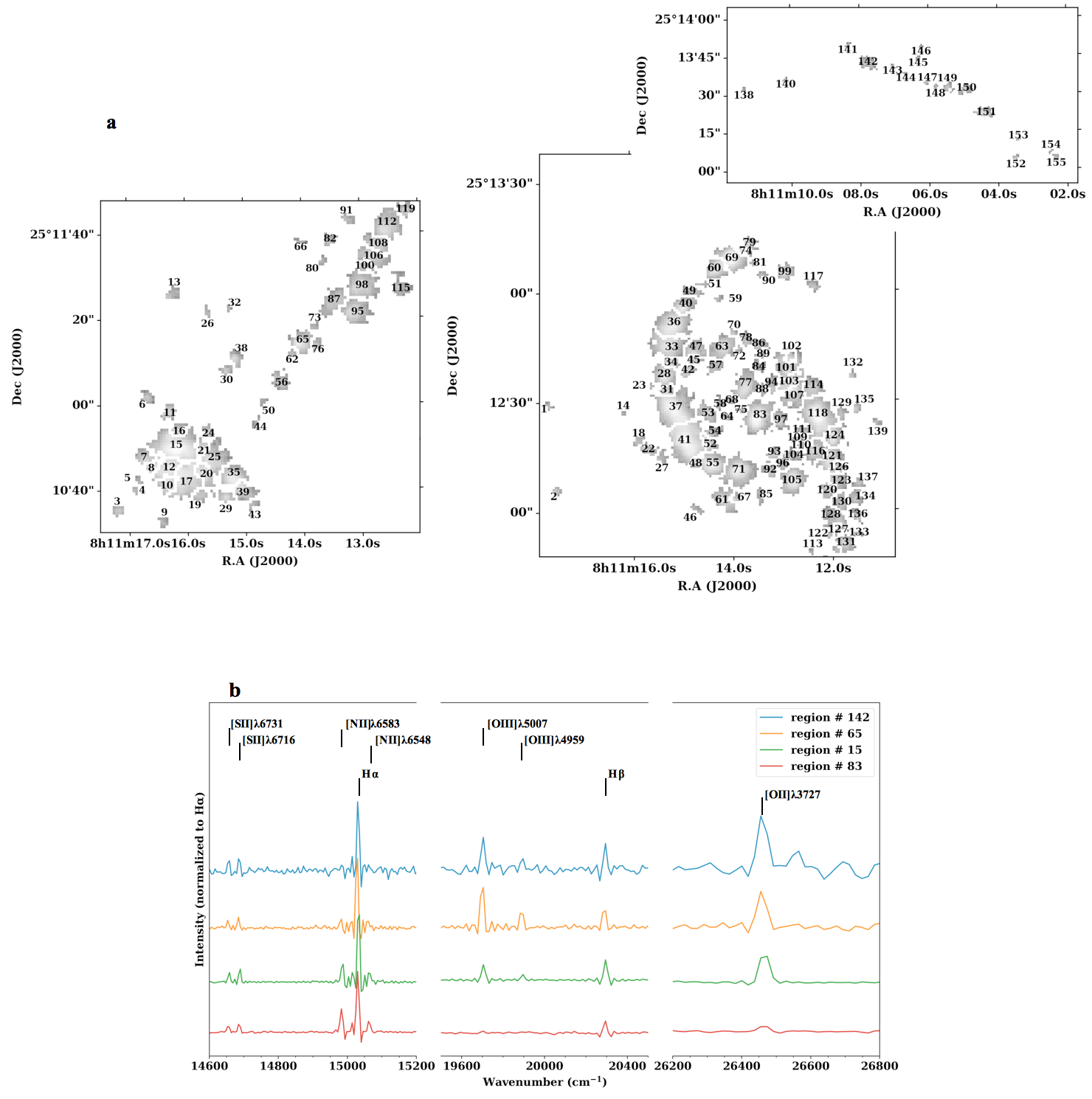}
    \caption{(a) HII region complexes detected in $\rm NGC\,2536$ and the bridge (left) and in $\rm NGC\,2535$ (right). (b) Spectra of regions \#143 (in the extended northern arm of $\rm NGC\,2535$), 65 (in the bridge), 15 (in the centre of $\rm NGC\,2536$) and 83 (in the centre of $\rm NGC\,2535$). Variations in the relative strength of high and low-ionization emission lines with respect to the Balmer lines hints for differences in metallicity.}
    \label{fig:HIIregions_figure}
\end{figure*}

We extract kinematic parameters of the interacting galaxies by modelling their velocity fields with DiskFit \citep[][]{Spekkens2007,Sellwood2010,Sellwood2015}. The code can fit axisymmetric and non-axisymmetric motions in two-dimensional velocity fields, as well as perform fits of outer disk symmetric warps. A model is fitted to regularly spaced, concentric ellipses at radii specified by the user, from which the kinematic parameters are derived assuming that the disk is flat (constant inclination and constant PA) and that the perturbation that drives the non-circular motions has a fixed principal axis. In our case, the separation between the ellipses were chosen to be two pixels to ensure that each ring contained a minimum of ten measured velocities at the lowest radii. The velocity model is given by the following expression :
\begin{equation}
    \begin{split}
    V_{\rm model} = V_{\rm sys} + 
    \sin{i} \big[\overline{V}_t\cos{\theta} - 
    V_{m,t}\cos{(m\theta_b)}\cos{\theta} \\
    - V_{m,r}\sin{(m\theta_b)}\sin{\theta}\big]    
    \end{split}
	\label{eq:velocitymodel}
\end{equation}
where $V_{\rm sys}$ is the systemic velocity, $\overline{V}_t$ is the circular velocity, $V_{m,t}$ and $V_{m,r}$ are the tangential and radial components of non-circular flows with harmonic order $m=1$ or $m=2$ in the disc plane, $\theta$ and $\theta_b$ are the azimuthal angles relative to the major axis and the non-circular flow axis, respectively, and $i$ is the disk inclination. If m = 1, the model describes a lopsided flow; if m = 2 the model is bi-symmetric, and describes a barred or elliptical flow. For an axisymmetric model, $V_{m,t}$ and $V_{m,r}$ are set to zero.

To avoid strong deviations from pure circular motions in our data, we only fit the inner parts of the discs of the two galaxies. For example, sampling data points on the velocity map at position $\sim$ [8h11m12s,$25\degree13^{\prime}00^{\prime\prime}$] would induce a gradient in projected velocities of $\sim$ 60\,km/s just across the northern spur of $\rm NGC\,2535$. A steep decrease in velocities would also occur for data points at approximately the same radius crossing the bridge. This behavior can be seen in the velocity profile of \citet{Amram1989}. Also, velocities along the tidal tails (extended arm and bridge) seem to remain constant over long patches, in agreement with the long-slit spectral observations of \citet{Zasov2019}. Results of the modelling are shown in Fig.~\ref{fig:Diskfitvelmod_figure}. $\rm NGC\,2535$ is best fitted by an axisymmetric model (Fig.~\ref{fig:Diskfitvelmod_figure}a). The residuals (Fig.~\ref{fig:Diskfitvelmod_figure}e) of the rotation-only model (Fig.~\ref{fig:Diskfitvelmod_figure}b) applied to $\rm NGC\,2536$ show clear indications of the presence of a non-axisymmetric flow. Adding a bar-like feature (Fig.~\ref{fig:Diskfitvelmod_figure}c) to the rotation-only model significantly lowers the residuals (Fig.~\ref{fig:Diskfitvelmod_figure}f). The choice of these models is also consistent with the structural components found in photometry modelling as discussed below. The best-fitting kinematic parameters are listed in Table~\ref{tab:table2}. Uncertainties on the models parameters were estimated by generating 100 bootstrap realizations of each velocity model \citep[][]{Sellwood2010}. The derived systemic velocities of the two galaxies are very close. Their centres have a projected separation of 102$^{\prime\prime}$ or 29 kpc for the adopted distance which transcribes to a deprojected separation of 30.5 kpc if $\rm NGC\,2536$ is in the plane of $\rm NGC\,2535$ disk. Fig.~\ref{fig:Diskfitvelmod_figure}g and Fig.~\ref{fig:Diskfitvelmod_figure}h show the radial velocity distribution of the best models. The projected velocity profiles of the models are in good agreement with those of \citet{Amram1989}. A velocity profile extending beyond the ocular ring in $\rm NGC\,2535$, using the same kinematic parameters derived here, is presented in Section~\ref{sec:section4} and compared to the radial velocity distribution of the primary galaxy in the simulation. The decline in the velocity profile of $\rm NGC\,2536$ occurs at the position of the bar and is likely associated with strong non-circular gas motions caused by the bar. However, we do not observe the change of sign beyond the 10$^{\prime\prime}$ radius reported by \citet{Zasov2019}. We performed non-linear least-squares fitting to the rotation curves following \citet{Courteau1997} arctan function modified to account for a possible increase/decline in the rotational velocity at high galactocentric radii \citep[][]{Drew2018}:
\begin{equation}
    \overline{V_t}(r) = \frac{2}{\pi} V_a\arctan\biggl(\frac{r}{r_t}\biggr)+cr
    \label{eq:rotation curve}
\end{equation}
where $V_a$ is an asymptotic velocity, $c$ is the outer slope and $r_t$ is a transition radius between the rising inner part and outer part. Fits to the rotation curves with this function are represented by dashed lines in Fig.~\ref{fig:Diskfitvelmod_figure}g and Fig.~\ref{fig:Diskfitvelmod_figure}h.

\begin{table}
 \caption{Best-fitting kinematic parameters. $x_c$ and $y_c$ are positions of the kinematic centre, $\Phi_{{\rm kin},d}$ is the PA of the kinematic major axis of the disk in the sky plane, $\Phi_{{\rm kin},b}$ is the PA of the major axis of the bar in the disk plane. For velocity fields, PAs correspond to the receding side of the disk. $V_a$, $r_t$ and $c$ are parameters describing fits to the rotation curves (see Eq.\eqref{eq:rotation curve}).}\label{tab:table2}
    \centering
    \begin{threeparttable}
      \small
      \begin{tabular}{ccc}
      \hline
      \small\makecell{ } & \small\makecell{NGC 2535} &
      \small\makecell{NGC 2536} \\
      \hline
      \small\makecell{$x_c$ \\} & \small 8h11m13.5s
      & \small 8h11m15.9s \\
      \small\makecell{$y_c$ \\} & 
      \small $+25\degree12'24.6''$
      & \small $+25\degree10'46.7''$ \\
      \small\makecell{$V_{\rm sys} [{\rm km\,s^{-1}}]$\\} 
      & \small $4124.7\pm1.4$
      & \small $4118.8\pm1.2$ \\
      \small\makecell{$i$ \\} 
      & \small $18.1\degree\pm9.9\degree$
      & \small $48.1\degree\pm4.9\degree$ \\
      \small\makecell{$\Phi_{{\rm kin},d}$ \\} 
      & \small $154.1\degree\pm3.0\degree$
      & \small $211.8\degree\pm4.5\degree$ \\
      \small\makecell{$\Phi_{{\rm kin},b}$ \\} & \small $\cdots$
      & \small $56.0\degree\pm9.8\degree$  \tnote{a}\\
      \\
       \small\makecell{$V_a [{\rm km\,s^{-1}}]$\\} 
      & \small $376.1\pm44.2$
      & \small $155.3\pm41.7$ \\
       \small\makecell{$r_t$ [arcsec]\\} 
      & \small $2.26\pm0.56$
      & \small $1.88\pm0.92$ \\
       \small\makecell{$c [{\rm km\,s^{-1}\,arcsec^{-1}}]$\\} 
      & \small $- 4.1\pm1.9$
      & \small $- 5.3\pm2.6$ \\
      \hline
      \end{tabular}
      \begin{tablenotes}
       \item[a] DiskFit computes $\Phi_{{\rm kin},b}$ in the disk plane so that Eq.\eqref{eq:velocitymodel} can apply. The corresponding value in the sky plane is $256.6\degree$ (see Eq.(6) of \citet{Spekkens2007} for conversion details) 
      \end{tablenotes}
    \end{threeparttable}
\end{table}

\begin{table}
    \caption{Best-fitting photometric parameters. $x_c$ and $y_c$ are positions of the image centre, $\Phi_{{\rm phot},d}$ is the PA of the major axis of the disk in the sky plane, $\Phi'_{{\rm phot},b}$ is the PA of the major axis of the bar in the sky plane. $\epsilon_{\rm bar/bulge}$ are apparent ellipticities of the bar/bulge components. $I_{e,\rm disk/bulge}$, $r_{e,\rm disk/bulge}$, and $n_{\rm disk/bulge}$ are parameters describing fits to the disk/bulge brightness profiles (see eq.\eqref{eq:bulgemodel}). $I_e$ brightness are in units of $10^{-17}\mathrm{erg\,s^{-1}\,cm^{-2}}$\AA$^{-1}\mathrm{arcsec^{-2}}$. The bottom three parameters are the percentages of light coming from the galaxy components.}\label{tab:table3}
    \centering
    \small
    \begin{tabular}{ccc}
    \hline
    \small\makecell{ } & \small\makecell{NGC 2535} &
    \small\makecell{NGC 2536} \\
    \hline
    \small\makecell{$x_c$ \\} & \small 8h11m13.5s
    & \small 8h11m15.9s \\
    \small\makecell{$y_c$ \\} & 
    \small $+25\degree12'24.6''$
    & \small $+25\degree10'46.8''$ \\
    \small\makecell{$i$ \\} 
    & \small $51.7\degree\pm2.2\degree$
    & \small $54.0\degree\pm0.1\degree$ \\
    \small\makecell{$\Phi_{{\rm phot},d}$ \\} 
    & \small $56.6\degree\pm2.6\degree$
    & \small $50.7\degree\pm0.8\degree$ \\
    \small\makecell{$\Phi'_{{\rm phot},b}$ \\} & \small $\cdots$
    & \small $92.1\degree\pm7.7\degree$ \\
     \small\makecell{$\epsilon_{\rm bar}$ \\} & \small $\cdots$
    & \small $0.26\pm0.02$ \\
     \small\makecell{$\epsilon_{\rm bulge}$ \\} 
    & \small $0.00\pm0.00$
    & \small $0.10\pm0.12$ \\
    \\
     \small\makecell{$I_{e,\rm bulge}$\\} 
    & \small $3.69\pm1.88$
    & \small $1.84\pm0.89$ \\
     \small\makecell{$I_{e,\rm disk}$\\} 
    & \small $0.17\pm0.03$
    & \small $0.19\pm0.02$ \\
     \small\makecell{$r_{e,\rm bulge}$ [arcsec]\\} 
    & \small $1.38\pm0.63$
    & \small $2.42\pm0.48$ \\
     \small\makecell{$r_{e,\rm disk}$ [arcsec]\\} 
    & \small $36.08\pm5.67$
    & \small $17.05\pm1.81$ \\
     \small\makecell{$n_{\rm bulge}$ \\} 
    & \small $1.05\pm0.61$
    & \small $0.46\pm0.17$ \\
     \small\makecell{$n_{\rm disk}$ \\} 
    & \small $1.59\pm0.15$
    & \small $1.50\pm0.06$ \\
    \\
     \small\makecell{Per cent disk \\} 
    & \small $93.94\pm3.15$
    & \small $67.53\pm8.20$ \\
     \small\makecell{Per cent bulge \\} 
    & \small $6.06\pm3.15$
    & \small $18.16\pm9.26$ \\
     \small\makecell{Per cent bar \\} 
    & \small $\cdots$
    & \small $14.31\pm7.06$ \\
    \hline
    \end{tabular}
    \end{table}

DiskFit can also be used to fit axisymmetric and non-axisymmetric models to images. We model the continuum map since it is smoother than the H$\alpha$ flux map. The photometry branch of the code can fit up to three components simultaneously : a disk, a bar and a bulge. The code determines the best-fitting values for ($x_c$,$y_c$), disk PA, disk ellipticity (and corresponding inclination), bar PA, bar ellipticity. The only component that has an assumed light profile is the bulge, which is parameterized by a Sérsic function:
\begin{equation}
    I(r) = I_e \mathrm{exp}\Bigg\{-b_n\bigg[\bigg(\frac{r}{r_e}\bigg)^{1/n}-1\bigg]\Bigg\}
	\label{eq:bulgemodel}
\end{equation}
where $I_e$ is the brightness at the effective radius $r_e$, $n$ is the Sérsic index. The factor $b_n$ is a function of the shape parameter, $n$, such that $\Gamma(2n)=2\gamma(2n,b_n)$, where $\Gamma$ is the gamma function and $\gamma$ is the incomplete gamma function. The bulge is spheroidal with an apparent ellipticity $\epsilon_{\rm bulge}$, with the disk plane being the plane of symmetry.

Fig.~\ref{fig:Diskfitphotmod2535_figure} and Fig.~\ref{fig:Diskfitphotmod2536_figure} show the best photometric models and residuals to the SN3 band continuum map together with the corresponding brightness profiles. The black dashed lines in the bottom panel represent non-linear least squares fits to disk brightness profiles by Eq.~\eqref{eq:bulgemodel}. The best fitting photometric parameters are listed in Table~\ref{tab:table3}. 100 bootstrap realizations of each photometric model were used to estimate the uncertainties on the photometric parameters. 

We fit a bulge+disk+bar model for $\rm NGC\,2536$. The resulting bar has the appearance of a ring while the enclosed bulge has a Sérsic index characteristic of bars ($n=0.5$). The bulge and the bar contribute up to 32\% of the total brightness of the model. The residuals are higher for star-forming knots in the bar and bulge. The photometric centre, inclination and P.As values are very close to those derived from the analysis of velocity maps. The stellar mass of the galaxy may be estimated from its $\rm(g-i)$ and $\rm(i-H)$ colours \citep[][]{Zibetti2009}. From the SIMBAD astronomical data base, $\rm(g-i)=0.75$ and $\rm(i-H)=2.0$ which results in $\rm M/L_H\sim0.25$. Considering the adopted distance of 59.2 Mpc, $\mathrm{L_H}=1.3\times10^{10}L_\odot$ or $\mathrm{M_\star}=3.2\times10^9M_\odot$. The total HI mass of Arp~82 is $2.4\times10^{10}M_\odot$, of which $2.4\times10^9M_\odot$ is associated to $\rm NGC\,2536$ \citep[][]{Kaufman1997}. The dynamical mass can be estimated as $M_{\rm dyn}(r)=(V^2+3.36\sigma^2r/r_e)\times r/G$ where $G$ is the gravitational constant, $V$ is the rotational velocity, $\sigma$ is the isotropic velocity dispersion \citep[][]{Drew2018}. \citet{Kaufman1997} report velocity dispersions of $\sim$ 30 km/s all over the system for the HI gas. We measure velocity dispersions of about the same value (see Fig.~\ref{fig:emissionlines_figure}) for the ionized gas in star forming knots with the highest SNR. Assuming $\sigma$\,=\,30 km/s and $V(r_e)=54\,\rm km/s$ (inferred from the parameterized rotation curve) we find a  dynamical mass of $6.5\times10^9M_\odot$ enclosed in one effective radius.

For $\rm NGC\,2535$, we fit a bulge+disk model to the data. The bulge has a Sérsic index $n_{\rm bulge}=1$, characteristic of disks. The bulge contributes only 6\% to the model total brightness. The ocular ring and the tidal arms stand out in the residuals image. For this galaxy, we infer $V(r_e)=210\,\rm km/s$, corresponding to a dynamical mass of $1.1\times10^{11}M_\odot$ within a radius $r=r_e$. However, the photometric inclination and PA of the disk are very different from the equivalent kinematic parameters values. If the inclination deduced from the image analysis is correct, $V(r_e)$ would be a factor $\sin51.7/\sin18.1=2.5$ less and the above estimated dynamical mass would be $2.5\times10^{10}\,M_\odot$. $\rm(g-i)$ and $\rm(i-H)$ colours from SIMBAD (1.20 and 2.23 respectively) suggest a $\rm M/L_H\sim0.25$ or $\mathrm{M_\star}=5.4\times10^{10}M_\odot$. Thus, the photometric inclination is certainly not correct as the computed dynamical mass would be less than the baryonic mass. This explains the very low dynamical mass, comparable to the HI content of the system, deduced form the H$\alpha$ Fabry-Perot observations by \citet{Amram1989} who found an inclination $i=43\degree\pm10\degree$ based on the ellipticity of the inner isophotes. \citet{Kaufman1997}, using the \citet{Persic1991} relation adopt an inclination $i=23\degree\pm4\degree$ to get velocities consistent with the B band luminosity of $\rm NGC\,2535$. The kinematic inclination is therefore favored. Hence, the galaxy disk and the eye-shaped ring with a major axis at $117\degree\pm12\degree$ all have intrinsic elliptical shapes. The misalignment between the kinematic and photometric P.As and the intrinsic ellipticity of the stellar disk are induced by the interaction as discussed in Section~\ref{sec:section4}.

\subsection{Properties of HII regions}
\label{sec:section32}
\subsubsection{Region detection and integrated spectra}
\begin{figure}
    \includegraphics[scale=0.27]{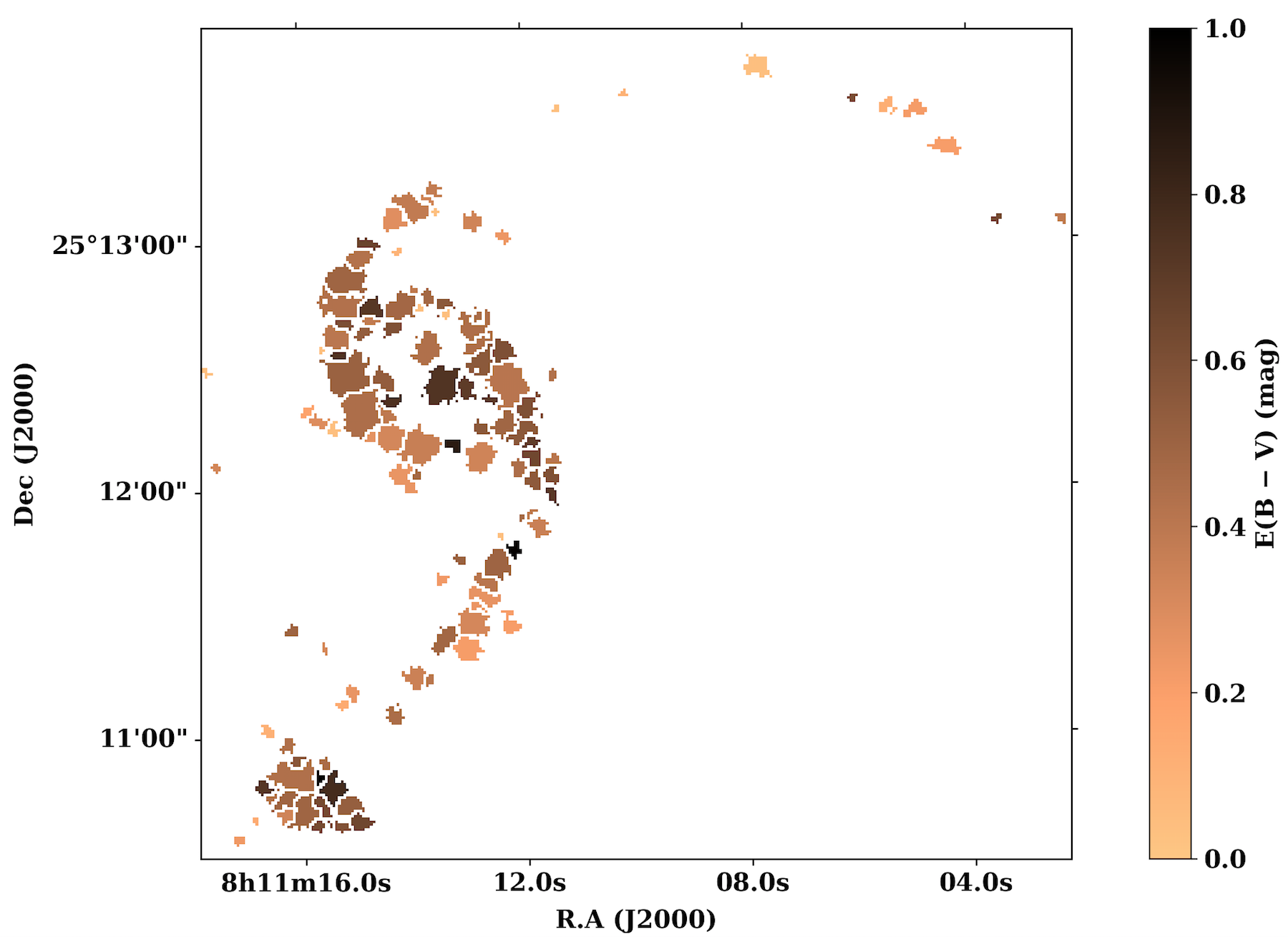}
    \caption{E(B\,\text{--}\,V) map.}
    \label{fig:extinction_figure}
\end{figure}

Gas phase metallicity is generally traced through oxygen abundance. Ionized gas oxygen abundances have been well calibrated on the basis of strong-line indicators for ionized regions associated with SF processes. We have selected HII region complexes using the automated HIIphot algorithm \citep[][]{Thilker2000,Thilker2002}. The procedure applied on the $\rm H\upalpha$ emission line and continuum maps detects 155 sources above the $10\sigma$ level. The selected HII region complexes are shown in Fig.~\ref{fig:HIIregions_figure}a. Positions and fluxes of the detected regions can be found in Table~\ref{tab:tableA1}.
Fluxes of the HII region complexes are then extracted from their sky subtracted integrated spectra in the same manner described in Section~\ref{sec:section2}, i.e, emission lines are fitted using sinc profiles for the SN1, SN2 data cubes and sincgauss profiles for the SN3 data cube. Visual inspection of the spectra do not show any significant underlying Balmer absorption. We, therefore, do not subtract any model stellar spectra from the extracted integrated spectra. Fig.~\ref{fig:HIIregions_figure}b illustrates spectra of some bright regions detected near the galaxy centres (regions \#15 and 83) and in the tidal tails (regions \#65 and 142). We corrected extracted fluxes for Galactic extinction and internal dust attenuation derived from the Balmer decrement $\rm H\upalpha/H\upbeta$. For the Galactic extinction, we assume the Cardelli extinction law \citep[][]{Cardelli1989} with $\rm R_v=3.1$ and $\rm A_v=0.117$, as provided by the NED astronomical database. For the internal dust attenuation, we assume a Calzetti law \citep[][]{Calzetti2001} and a theoretical H$\upalpha$/H$\upbeta$\,=\,2.86 for the Case B recombination at T\,=\,10\,000\,K, $\rm n_e=100\,cm^{-3}$ \citep[][]{Osterbrock&Ferland2006}. We discard 40 regions with H$\upalpha$ or H$\upbeta$ fluxes with a SNR $\leq$ 3. The resulting color excess $\rm E(B-V)$ map is shown in Fig.~\ref{fig:extinction_figure} with a median value of $\sim0.44$.

\begin{figure}
    \centering
    \includegraphics[scale=0.35]{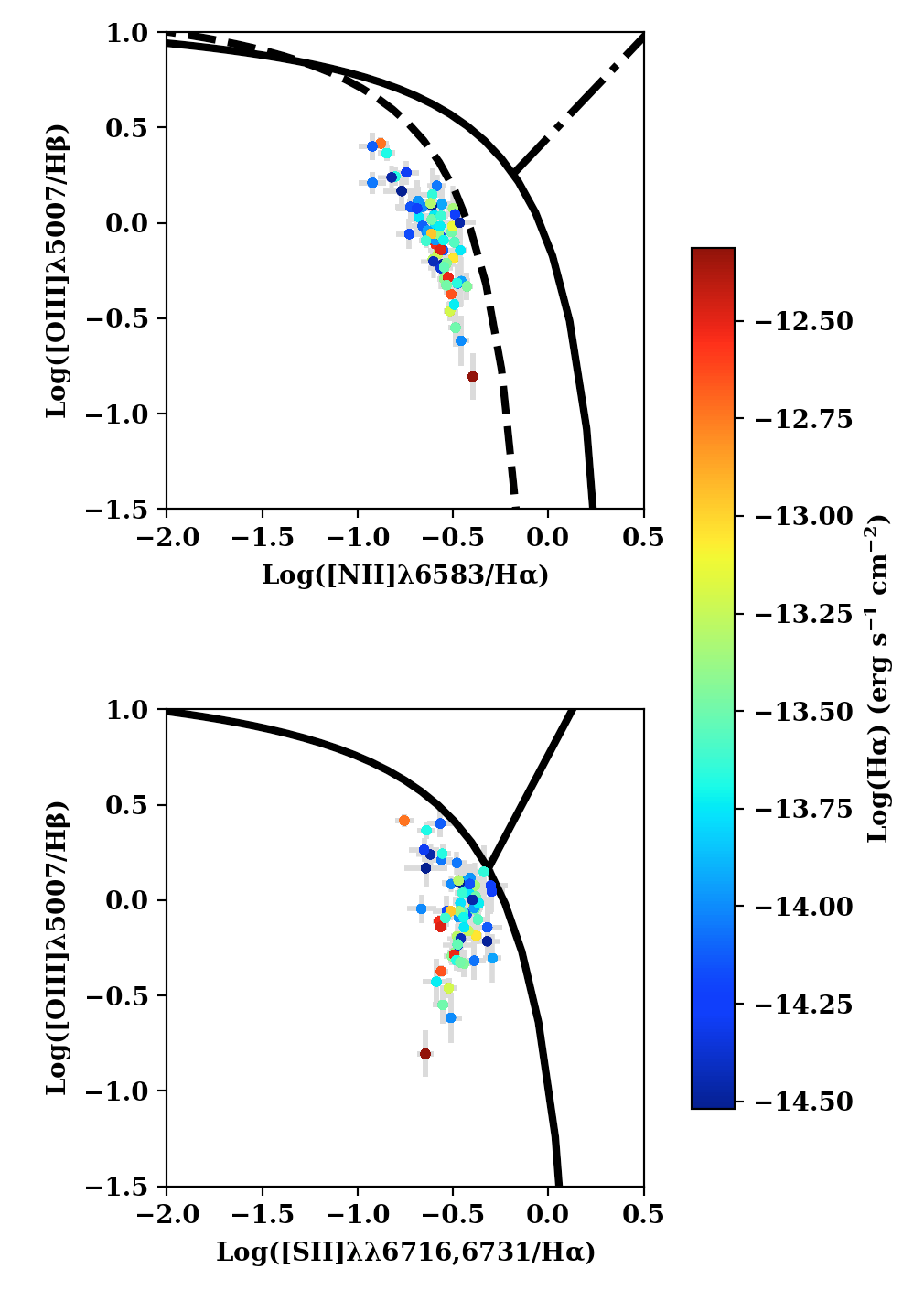}
    \caption{BPT diagrams showing excitation properties of detected HII regions (see text). Demarcation lines from \citet{Kewley2001}(solid lines), \citet{Kauffmann2003}(dashed line), \citet{Schawinski2007}(dash-dotted line) are plotted and separate star forming regions, high ionization and low ionization emission regions. Color denotes extinction corrected H$\upalpha$ fluxes.}
    \label{fig:BPT_figure}
\end{figure}

Selecting regions solely based on their brightness, however, does not provide information on the ionization mechanisms. In order to exclude sources other than SF processes, we need to inspect their locations on the BPT diagrams \citep[][]{BPT1981}. To this purpose, we only consider regions with a SNR $>$ 3 for all emission lines before doing the diagnostic. This process further excludes 45 regions. Thus, of the 155 regions detected with HIIphot, we finally only keep 70. The [NII]$\uplambda$6583/H$\upalpha$ vs. [OIII]$\uplambda$5007/H$\upbeta$ and [SII]$\uplambda\uplambda$6716,6731/H$\upalpha$ vs. [OIII]$\uplambda$5007/H$\upbeta$ diagnostic line ratios are shown in Fig.\ref{fig:BPT_figure}. All selected regions lie leftward (within uncertainties) of the demarcation lines in the BPT diagrams (Fig.~\ref{fig:BPT_figure}).

\begin{figure*}
    \includegraphics[width=\textwidth]{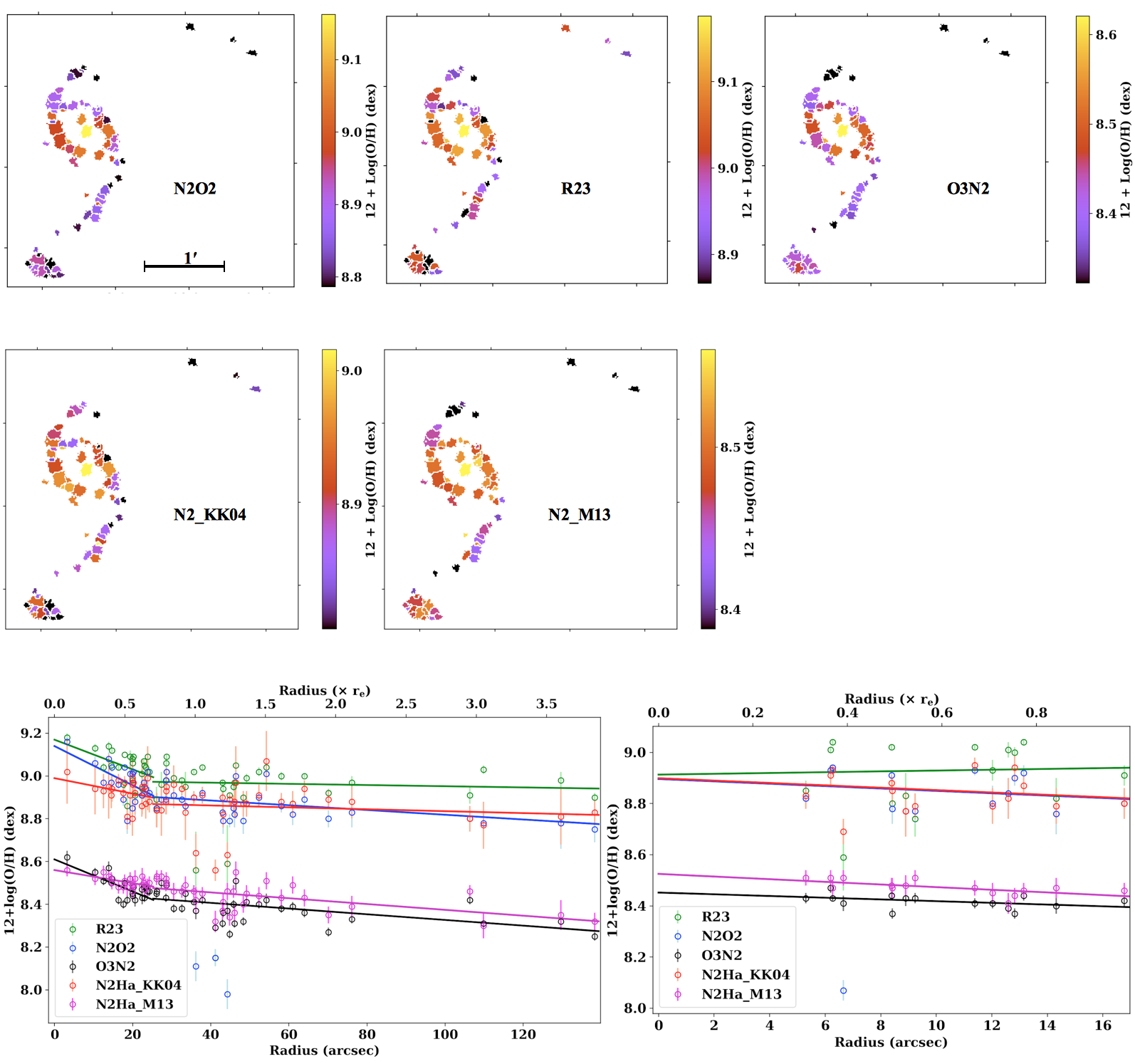}
    \caption{Abundance maps (top and middle panels) and abundance profiles (lower panel) for $\rm NGC\,2535$ and $\rm NGC\,2536$. The maps and profiles are obtained for the N2O2, R23 and O3N2, N2Ha$\_$KK04 and N2Ha$\_$M13 calibrations. Linear fits (solid lines) to the data points are performed ignoring outliers (see text).}
    \label{fig:abundance_figure}
\end{figure*}

\subsubsection{Oxygen abundance determination}
To evaluate gas phase metallicity, we chose five different calibrators widely used in the literature, and compare the derived abundances to avoid systematic biases that may arise from the adopted extinction, diffuse ionized gas (DIG) contribution, low ionization emission regions (LIERs), ionization parameter or relative elemental abundances variations (e.g., N/O variations when using indicators based on [NII] emission lines to derive metallicity). Those indicators are N2Ha \citep[][]{Kobulnicky2004, Marino2013}, based on $\rm([NII]\uplambda6583/H\upalpha)$, O3N2 \citep[][]{Marino2013}, based on ([OIII]$\uplambda$5007/H$\upbeta$)/([NII]$\uplambda$6583/H$\upalpha$) line ratio, N2O2 \citep[][]{Kewley2002,Bresolin2007}, based on [NII]$\uplambda$6583/[OII]$\uplambda\uplambda$3727,3729 line ratio and R23 \citep[][]{Kobulnicky2004}, based on $\rm([OII]\uplambda\uplambda3727,3729+[OIII]\uplambda5007)/H\upbeta$ line ratio. N2Ha is not sensitive to reddening correction or flux calibration as the involved lines are very close in wavelength but is particularly sensitive to the ionization parameter and to the presence of strong shock excitation or DIG/LIERs contribution. The indicator scales linearly with nitrogen abundance at low metallicities ($\rm N2Ha < -0.8$) but saturates at high metallicities \citep[][]{Kewley2002}. \citet{Kumari2019} have shown that the O3N2 indicator is not significantly affected by DIG/LIERs. The indicator also has the advantage to vary monotonically with metallicity. However, the diagnostic ratio is affected by variations in the ionization parameter since [NII] and [OIII] originate from ions with a large difference in ionization potential. N2O2, on the other hand, is ionization parameter independent and is a good diagnostic, but only for metallicities above half solar \citep[][]{Kewley2002}. The R23 parameter has a local maximum and is often used in conjunction with another indicator like N2O2 and an ionization parameter estimator like the O3O2 (\,log([OIII]$\uplambda$5007/[OII]$\uplambda\uplambda$3727,3729)) and gives reliable estimates of oxygen abundances at low metallicities. For O3N2, we used the empirical calibration of \citet{Marino2013}, valid for $\rm-1<\log[([OIII]\uplambda5007/H\upbeta)/([NII]\uplambda6583/H\upalpha)]<1.7$ (all HII region complexes analyzed here fall in this range). For the N2O2 diagnostic, we used the theoretical calibration provided by \citet{Kewley2002} for regions with $\rm 12+\log(O/H)>8.6$ or $\rm\log([NII]\uplambda6583/[OII]\uplambda\uplambda3727,3729)>-0.97$. For $\rm\log([NII]\uplambda6583/[OII]\uplambda\uplambda3727,3729)<-0.97$ we used the empirical \citet{Bresolin2007} prescription. For R23, we used the iterative procedure described in \citet{Kobulnicky2004} to derive (O/H) abundances and the ionization parameter simultaneously. For N2Ha, we adopt the parameterization given by \citet{Kobulnicky2004} (N2Ha$\_$KK04 hereafter) and use the final ionization parameter map obtained by the R23 calibration to derive (O/H). We also use the \citet{Marino2013} empirical calibration (N2Ha$\_$M13) for the same indicator. (O/H) abundance maps derived using the five different methods are reproduced in Fig.~\ref{fig:abundance_figure} together with the (O/H) radial distribution in the disks of $\rm NGC\,2535$ and $\rm NGC\,2536$. Deprojected galactocentric distances were computed taking into account the kinematical PAs and inclinations found in Section \ref{sec:section31}. Absolute metallicity estimates from the five calibrations are different with systematically lower values obtained for the empirical calibrations (O3N2 and N2Ha$\_$M13). This discrepancy is expected as theoretical calibrations usually lead to higher (O/H) values than empirical ones \citep{KewleyEllison2008}\footnote{However, we note that, on the one hand, some tailor-made models for low metallicity objects \citep[e.g.,][]{PerezMontero2010} and some recent state-of-the-art models that assume empirical laws between O/H, ionization parameter and N/O \citep[e.g.,][]{PerezMontero2014} give chemical abundances in good agreement with those derived from the direct method for high metallicity objects. On the other hand, some recent empirical calibrations \citep[e.g.,][]{Pilyugin2016} provide estimates of abundances also compatible with those from the direct method.}. This also explains the low values found for region \#21 in NGC\,2536 and regions \#99, 119, 134 in NGC\,2535 where the empirical calibration of \citet{Bresolin2007} was used for the N2O2 indicator. The corresponding R23 values are also low as N2O2 values serve as initial guesses for this calibrator. We find an offset of $\sim0.5\,\rm dex$ between the theoretical and empirical calibrations. The O3N2 values are less scattered in the radius vs. (O/H) diagrams. Central values of oxygen abundance found using the O3N2 calibration are in agreement with those derived by \citet{Zasov2019} who accounted for underlying stellar absorption in measuring emission line fluxes. Had the underlying H$\upbeta$ absorption line been important and not detected in our spectra due to the low resolution of the SN2 cube, we would have underestimated the H$\upbeta$ fluxes. As a consequence, we would have overestimated the internal extinction and significantly underestimated central oxygen abundances.

\subsubsection{Galactocentric abundance gradients}
Spatial distribution of oxygen abundance in the disk of $\rm NGC\,2535$ exhibits two slopes for all the methods used. The break occurs at $\sim25^{\prime\prime}$, just beyond the ocular ring. Indeed, the tidal tails have constant metallicity over long patches in the abundance maps. This is in agreement with the nearly constant abundance along the tidal bridge observed by \citet{Zasov2019}. Linear least-squares fit give slopes of $0.04\pm0.02$, $-0.05\pm0.02$, $-0.04\pm0.01$, $-0.02\pm0.02$, $-0.05\pm0.01$ dex/$r_e$ for N2O2, R23, O3N2, N2Ha$\_$KK04, N2Ha$\_$M13 calibrations respectively. Regions \#21, 99, 119, 134 were considered as outliers and were not used in the fits since they depart considerably from the general trend observed for the other regions for the N2O2 and R23 calibrators. Inside the ocular ring, the slope is steeper. We find gradients of $-0.33\pm0.11$, $-0.25\pm0.10$, $-0.27\pm0.06$, $-0.13\pm0.08$, $-0.11\pm0.03\,{\rm dex}/r_e$ with central (O/H) abundances of $9.14\pm0.05$, $9.17\pm0.05$, $8.61\pm0.03\,\rm dex$ for N2O2, R23, O3N2, N2Ha$\_$KK04, N2Ha$\_$M13 calibrations respectively. The inner slope is a bit less steep for the N2Ha calibrator since it saturates at high metallicities but the break in the O/H distribution is clear. The slopes derived are consistent within errors. The lower (O/H) values in the tidal tails suggest that these features are made of gas expelled from the disk periphery. In $\rm NGC\,2536$, the oxygen abundance distribution can be fitted with a single slope. We find gradients of $-0.08\pm0.09$, $0.02\pm0.15$, $-0.05\pm0.03$, $-0.08\pm0.08$,
$-0.08\pm0.03\,{\rm dex}/r_e$ with central abundances of $8.89\pm0.06$, $8.91\pm0.09$, $8.45\pm0.02$, $8.89\pm0.05$, $8.52\pm0.02\,\rm dex$ for N2O2, R23, O3N2, N2Ha$\_$KK04, N2Ha$\_$M13 calibrations respectively. The slopes derived for $\rm NGC\,2536$ are also consistent within errors. It is interesting to note that central metallicity in $\rm NGC\,2536$ is higher compared to the metallicity in the bridge, suggesting that the gas in the satellite galaxy has not yet mixed with the gas in the tidal bridge. \citet{Sanchez2014} showed that, when using the O3N2 indicator, disk galaxies in the local universe present a common gradient in the oxygen abundance of $\sim-0.1\,{\rm dex}/r_e$ up to $\sim2$ effective radii beyond which the distribution flattens out. The characteristic gradient is independent of morphology, presence or absence of bars, absolute magnitude, or stellar mass. Only interacting galaxies in their sample show clear deviations from the common gradient with an average slope of $-0.05\,{\rm dex}/r_e$. This is the case with the average slope derived for $\rm NGC\,2536$. For $\rm NGC\,2535$, only the outer slope is significantly shallow to this standard. Flat slopes in tidal debris have also been observed in other interacting systems \citep[][]{Weilbacher2003,Torres-Flores2014}. Since a metallicity gradient is expected at radii $< 2 \rm r_e$ for non-interacting disk galaxies, the flat outer slope hints at some mixing in the tidal debris or tidal stretching induced by the interaction. At this early stage of the interaction, tidal stretching seems more likely to cause the flattening.

\begin{table*}
    \caption{Initial properties of the simulated galaxies. DM, gas, star labels refer to the dark matter, gaseous, stellar components respectively. M, $\ell$ and N represent the total mass, scale length and total number of particles respectively.}\label{tab:initial1}
    \centering
    \small
    \begin{tabular}{lcccccrrr}
    \hline
    \smallskip
    Galaxy & $M_{\rm DM} [{\rm 10^9M_\odot}]$ & $M_{\rm gas} [{\rm 10^9M_\odot}]$ & 
    $M_{\rm star} [{\rm 10^9M_\odot}]$ & $\ell_{\rm gas}\ [{\rm kpc}]$ & $\ell_{\rm star}\ [{\rm kpc}]$ & $N_{\rm DM}$ & $N_{\rm gas}$
    & $N_{\rm star}$\\
    \hline
    Gal1 & 3154.265 & 32.304 & 51.504 & 6.0 & 3.0 & 630848 & 129216 & 206016 \\
    Gal2 &  492.853 & 2.692 & 4.292 & 2.0 & 1.0 & 98570 & 10768 & 17168\\
    \hline
    \end{tabular}
    \end{table*}

\begin{table}
    \caption{Initial conditions}\label{tab:initial2}
    \centering
    \small
    \begin{tabular}{lr}
    \hline
    ${\bf R}_1\, [{\rm kpc}]$ & (0,0,0)\\
    ${\bf R}_2\, [{\rm kpc}]$ & (74,20,8)\\
    ${\bf V}_1\, [{\rm km\,s^{-1}}]$ & (0,0,0)\\
    ${\bf V}_2\, [{\rm km\,s^{-1}}]$ & ($-$120,170,13)\\
    $(\theta,\phi)_1$ & (342,0)\\
    $(\theta,\phi)_2$ & (48,270)\\
    \hline
    \end{tabular}
    \end{table}

\subsubsection{Nucleus and star formation rate}
We do not detect any signature of AGN in our data. Inspection of the velocity dispersion map in Fig.~\ref{fig:emissionlines_figure} shows that a maximum of $\sim 45$ km/s is reached in the centre of $\rm NGC\,2535$. This value is very low compared to velocity dispersion of a few hundreds km/s observed in narrow line regions (NLRs) of active galaxies. It is even lower if compared to the widths of broad line regions. Extended NLRs would have similar widths but emission lines from the integrated spectra do not suggest the presence of multiple components associated to AGN-driven outflows/inflows and the diagnostic line ratios of Fig.~\ref{fig:BPT_figure} do not show any evidence of an AGN source. Although AGNs may be optically elusive, NIR spectra obtained by \citet{Lee2012} do not show any AGN signature either. The total H$\upalpha$ luminosity of the region complexes is $1.5 \times 10^{42}$ erg\,s$^{-1}$ for the adopted distance. The brightest regions are located at the centres of the galaxies, at the apices of the ocular and at the tip of the north spiral arm of NGC\,2535. Half of the total H$\upalpha$ luminosity is contributed by regions in the ocular. The proportion rises to seventy percent when accounting the spiral arms attached to the eye-shaped ring whereas the luminosity in the tidal tails amounts to five percent. Assuming that H$\upalpha$ luminosity is totally due to SF, we compute a global SFR of 8.1 $M_\odot$ yr$^{-1}$ for the Arp 82 system using the \citet{Murphy2011} calibration. The SFR would be 12.1 $M_\odot$ yr$^{-1}$ were it computed using \citet{Kennicutt1998ARA&A} calibration due to differences in stellar initial mass functions (IMF) and stellar population models (instantaneous vs. continuous SF) assumptions. The total infrared luminosity ($\rm L_{IR} = 1.1 \times 10^{44}$ erg\,s$^{-1}$) suggests a SFR measurement consistent with that traced by the H$\upalpha$ luminosity when considering the \citep[][]{Buat1996} relation valid for galaxies of type Sb and later. The derived SFR and total stellar mass (previous section) place the system on the local blue SDSS galaxies sequence in the SFR-M$_\star$ diagram of \citet{Elbaz2007} (see their Fig.~18 and Eq.5). Hence, the SF activity of Arp 82 is still comparable to typical star forming galaxies at $z=0$ with the same stellar mass.

\section{NUMERICAL MODEL}
\label{sec:section4}
\subsection{Simulation code}
\label{sec:section41}
To simulate the collision between the two galaxies, we used the galactic chemodynamical evolution code GCD+ \citep[][]{Kawata2003,Kawata2013}. GCD+ is a 3D tree N-body/SPH code that incorporates self-gravity, hydrodynamics, radiative cooling, star formation, supernova feedback, metal enrichment, and metal diffusion. Star formation occurs whenever the local number density of gas is greater than a threshold $\mathrm{n_{th}}$, the gas velocity field is convergent, and the gas is Jeans unstable following the Schmidt law:
\begin{equation}
    \frac{d\rho_*}{dt} = \frac{d\rho_g}{dt} = \frac{C_*\rho_g}{t_{g}},
	\label{eq:SF}
\end{equation}
where $\rho_{*}$ and $\rho_{g}$ are the stellar and gas mass density, respectively, $t_{g}$ the dynamical time and $C_{*}$ a dimensionless star formation efficiency. We use $\rm n_{th}=0.1\,{\rm cm}^{-3}$ and $C_*=0.012$. Dark matter, gas, and stellar components are represented by particles. Stellar and gas particles have equal masses. Each stellar particle represents a population of stars born at the same time with stellar masses distributed according to the \citet{Salpeter1955} IMF. The stellar wind (SW) energy output $E_{\rm SW}=10^{36}{\rm erg\,s}^{-1}$ and the supernova (SN) energy output $E_{\rm SN}=10^{51}{\rm erg}$ with only 10\% contributing to feedback, the rest being radiated away.

Since the algorithm does not include an actual treatment of ionization, we assume that ionized gas is traced by gas particles with high density and temperature (average $\mathrm{n_{H}}>1\,{\rm cm}^{-3}$, $T>420\,\rm K$). The assumption is made considering that in the Galaxy, the hot and diffuse ionized gas have densities well below $1\,\rm cm^{-3}$ and the mass weighted average temperature of the atomic gas and ionized gas by HII regions is $\sim$ 420 K.

\subsection{Initial conditions}
\label{sec:section42}
The simulation is run assuming a $\Lambda$CDM standard cosmology with $h=0.73$, $\mathrm{\Omega_0}=0.266$, $\mathrm{\Omega_b}=0.044$, $\mathrm{\lambda_0}=0.734$. Each galaxy is composed of a dark matter halo with a NFW profile \citep[][]{NFW1996} with a concentration parameter $c=20$, and a galactic disk with exponential density profile :
\begin{equation}
    \rho = \frac{M}{4 \pi \zeta \ell^2} \mathrm{sech}^2 \left( \frac{z}{\zeta} \right) \mathrm{exp}\left(-\frac{R}{\ell} \right),
	\label{eq:expprofile}
\end{equation}
where $R$ and $z$ are the radial and vertical coordinates, $\ell$ is the scale length, $\zeta$ is the scale height (set to $\ell$/8), $M$ is the mass of the stellar/gaseous component. The initial masses, %characteristic 
scale lengths, and number of particles for the simulated galaxies are given in Table~\ref{tab:initial1}. The stellar and gaseous disks of the primary and satellite galaxies have initial central iron abundances $\rm[Fe/H]$ of 0.2 and $0\,\rm dex$, respectively, with radial gradients of $-0.03$ and $-0.06\,\rm dex/kpc$ respectively. $\alpha$-elements are initially only present in the stellar component with abundances given by:
\begin{equation}
    \rm[\alpha/Fe] = -0.16[Fe/H].
	\label{eq:abundprofile}
\end{equation} 
The metallicity of each particle is modified by adding a Gaussian scatter of $0.02\,\rm dex$ to create a local dispersion. Stellar particles are assigned an age following an age-metallicity relation $\rm[Fe/H]=-0.04\times age(Gyr)$.
The simulated interacting galaxies both rotate counterclockwise and the encounter is prograde-retrograde with initial positions, velocities and inclination angles given in Table~\ref{tab:initial2}.

Several simulations were run in order to reproduce the morphology, kinematics, metallicity distribution, distance, global SFR observed in Arp 82. The best match simulation was found using the parameters described above and was run for $1\,\rm Gyr$. The satellite galaxy is on a bound orbit with initial positions and velocities chosen to reproduce the observed distance at the best match time. At this time, the kinematics and morphologies of the galaxies, in particular tidal tails induced by the interaction, are influenced by the choice of the initial scale lengths and initial inclinations. Initial metallicity and radial gradients were chosen to reproduce the observed metallicity distribution (derived using the N2O2 calibrator). Initial masses were chosen to reproduce the observed stellar and gaseous masses at the best match time.

\subsection{Simulations}
\label{sec:section43}

\begin{figure}
    \includegraphics[scale=0.25,center]{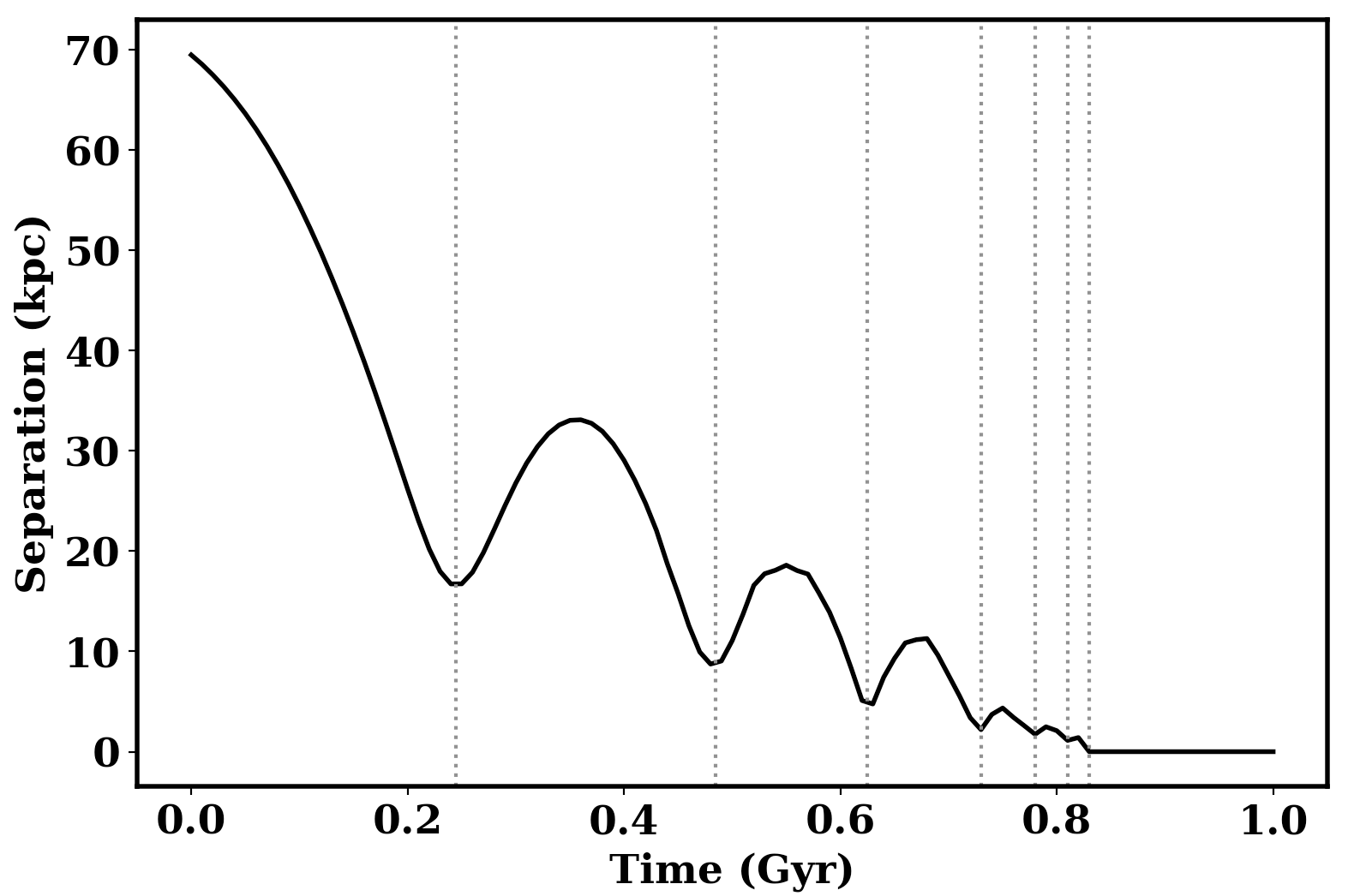}
    \caption{Temporal evolution of the separation between the centres of mass of the simulated interacting galaxies. Dotted lines identify instants at pericentre passages.}
    \label{fig:separation_figure}
\end{figure}

\begin{figure*}
    \includegraphics[scale=0.72]{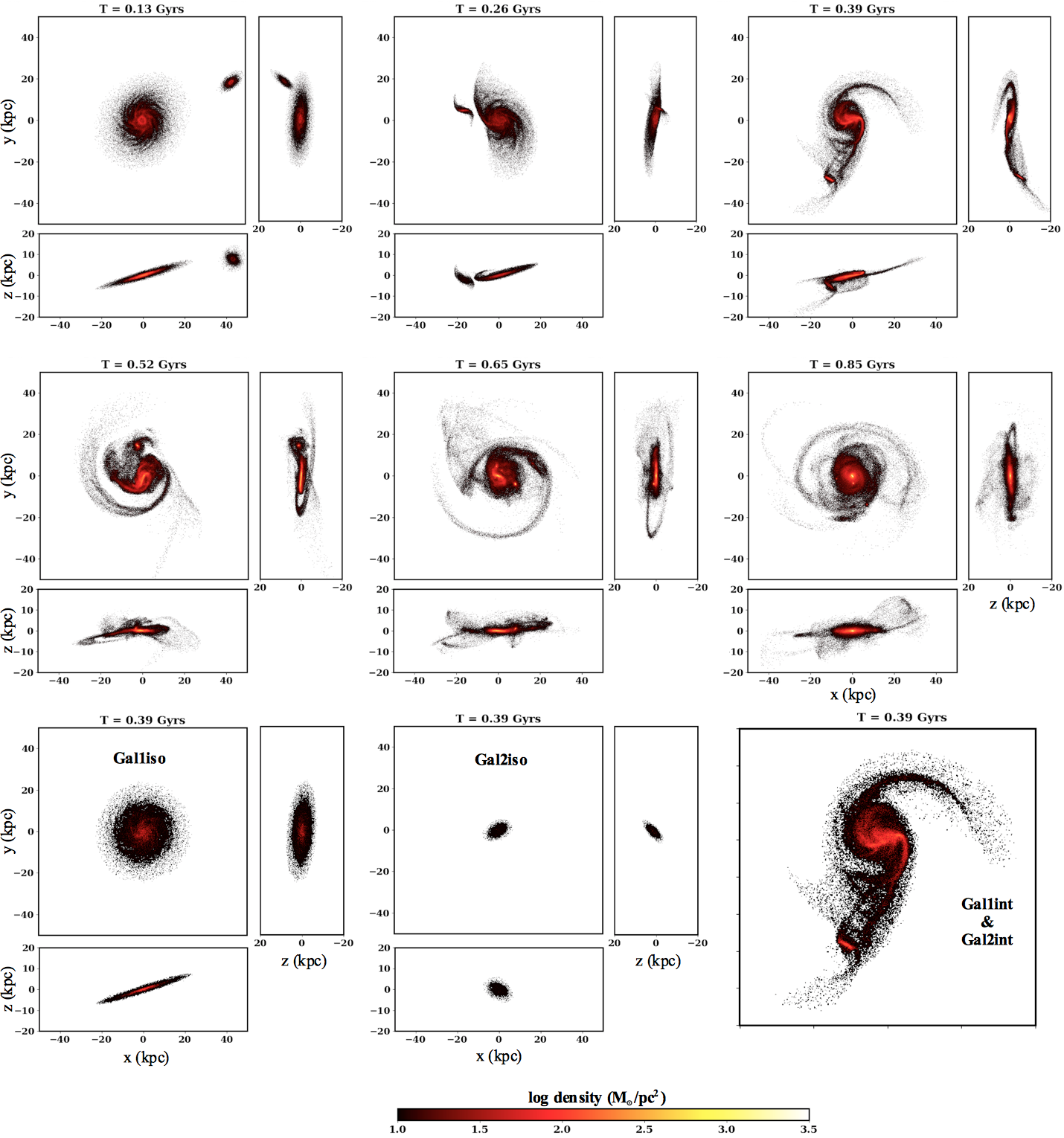}
    \caption{Temporal evolution of the stellar component of the simulated galaxies. Snapshots in the top and middle panels represent surface density maps of simulated interacting galaxies at selected times in the $xy$, $yz$ and $xz$ planes. The $xy$~plane is the plane of the sky. The best match configuration to observations occurs at $t=0.39\,\rm Gyr$ ; it is best shown in the xy plane in the lowest rightmost snapshot. The lowest left and middle snapshots represent stellar surface density maps of the galaxies evolved in isolation at $t=0.39\,\rm Gyr.$}
    \label{fig:simustars_figure}
\end{figure*}

\begin{figure*}
    \includegraphics[scale=0.80]{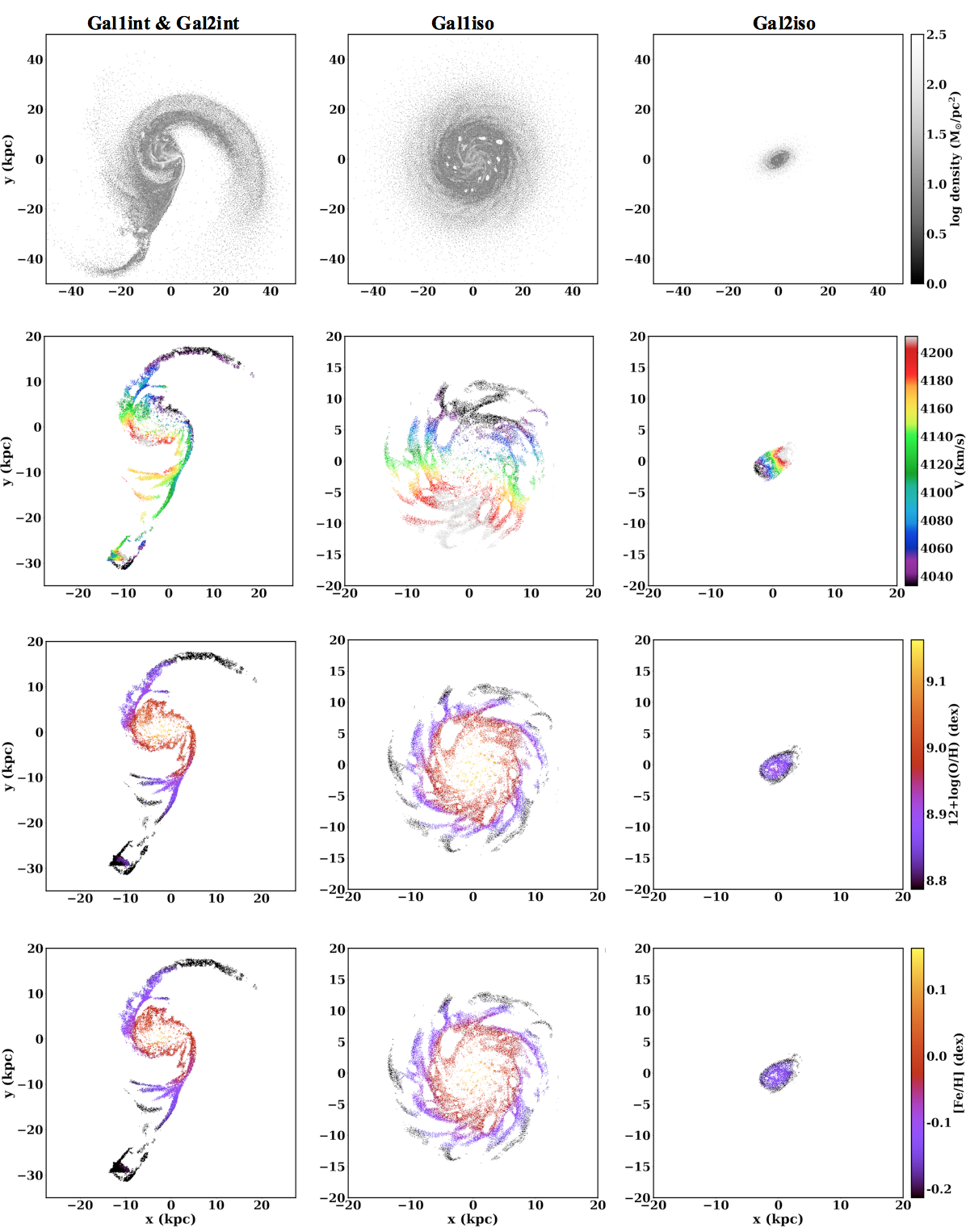}
    \caption{Gas surface density maps (1st row), ionized gas velocity maps (2nd row), oxygen abundance maps (3rd row) and iron abundance maps (4th row) for the simulated galaxies at $t=0.39\,\rm Gyr$.}
    \label{fig:simugas_figure}
\end{figure*}

\begin{figure}
    \includegraphics[width=\columnwidth]{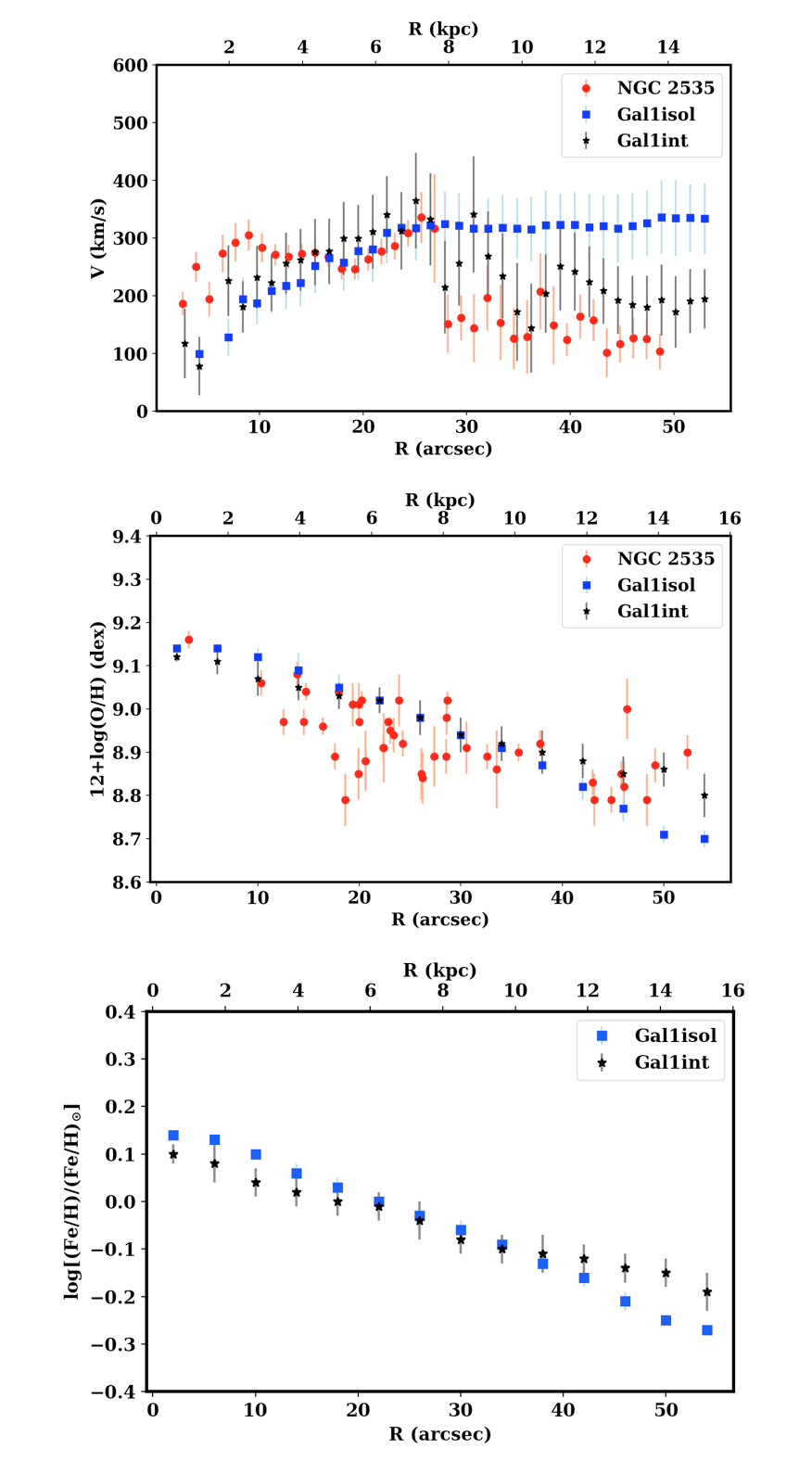}
    \caption{Velocity profiles (upper panel) and oxygen abundance profiles (middle panel) of Gal1int, Gal1iso, at $t=0.39\,\rm Gyr$, and $\rm NGC\,2535$. The bottom panel shows the corresponding iron abundance profiles of Gal1int and Gal2int.}
    \label{fig:VOHprof_figure}
\end{figure}

\begin{figure*}
    \includegraphics[scale=0.47]{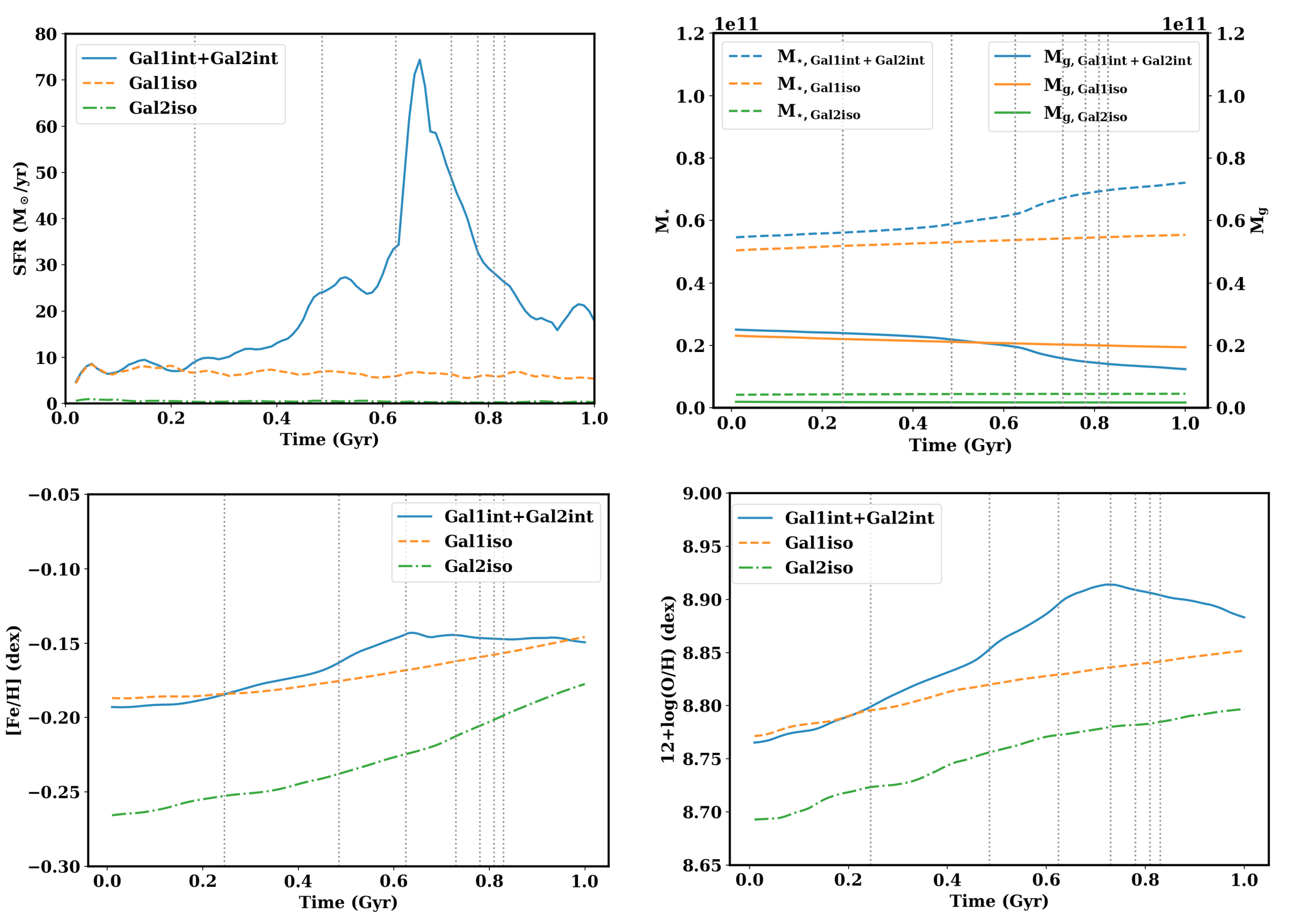}
    \caption{Temporal evolution of global star formation, stellar mass, gas mass, average Fe/H and O/H in Gal1int+Gal2int, Gal1iso and Gal2iso.}
    \label{fig:globalevolution_figure}
\end{figure*}

To investigate the effects of the interaction, we also ran simulations of the same galaxies with the same parameters but evolved in isolation. In the following, we refer to the primary and secondary galaxies as Gal1int and Gal2int, if simulated in interaction, and Gal1iso and Gal2iso when they are evolved in isolation.

Gal1int and Gal2int are initially separated by $\sim70\,\rm kpc$ and experience several close passages in their encounter before merging at $t=0.83\,\rm Gyr$ as shown in Fig.~\ref{fig:separation_figure}. The best match to the observations occurs at $t=0.39\,\rm Gyr$, just past the first apocentre, when the distance ($30.6\,\rm kpc$) between Gal1int and Gal2int very well matches the measured separation between $\rm NGC\,2535$ and $\rm NGC\,2536$. We thus identify $t=0.39\,\rm Gyr$ as ``the present''.

\subsubsection{Formation and evolution of tidal features}
Fig.~\ref{fig:simustars_figure} illustrates the time evolution of the configuration of the stellar particles in three orthogonal ($xy$, $yz$, $xz$) projections. The $xy$~plane is adopted as the plane of the sky and all particles are rotated by $15\degree$ in this plane around the origin after the run. The interaction strengthens spiral arms in Gal1int and tidal forces elongate the stellar disks as Gal1int and Gal2int approach the first pericentre passage. The elongation of Gal1int in the direction joining the centres of the interacting galaxies at first pericentre passage explains the intrinsic ellipticity and photometric P.A. observed for the stellar disk of $\rm NGC\,2535$. The distortion increases as the interaction progresses, with the outer parts of the disks being swept up in the tidal tails. The top right panel
of Fig.~\ref{fig:simustars_figure} and its close up view in the $xy$ plane shown in the bottom right panel show the best match stellar configuration to the observed interacting galaxies and happens when the satellite galaxy is past the first pericentre, half way to the second closest passage. Besides the strong tidal arms, Gal1int has an ocular ring and spurs which result from spiral arms existing at earlier times. The spur that crosses the bridge between Gal1int and Gal2int and the spur above the ocular ring, together with the bases of the tidal arms delineates an elliptical arc. The elliptical arc, as described by \citet{Kaufman1997}, was not detected in their B-band and I-band images despite being similar in shape and orientation to the outer isophotes of the stellar disk of $\rm NGC\,2535$ in those images. This lead the authors to suggest that this structure was a purely gaseous feature resulting from a wake in the gas caused by the passage of the companion within or close to the extended HI envelope of $\rm NGC\,2535$. The stellar counterpart, as reproduced by our simulations, can be seen in the deep images of Fig.~\ref{fig:deepimages_figure} ; the curved spurs of star-forming regions crossing the tidal tails are the brightest parts of the arc at its eastern and western boundaries. The structure was also detected in the UV images of \citet{Hancock2007}. The simulated isolated galaxies at $t=0.39\,\rm Gyr$ are shown in the bottom left and bottom middle panels of Fig.~\ref{fig:simustars_figure}. The regular circular disks of Gal1iso and Gal2iso contrast those of Gal1int and Gal2int and merely change their morphologies during the first $1\,\rm Gyr$ of evolution. The central densities of Gal1int and Gal2int increase significantly past the second pericentre passage at $t=0.49\,\rm Gyr$ as their separation shortens. The simulated interacting galaxies merge at $t=0.83\,\rm Gyr$ to form a puffy remnant with shells and extended tidal arms. Although the best match to the observations happens at $t=0.39\,\rm Gyr$ Gyr, we point out that Gal1int's extended tidal arm, on the anti-companion side, is not as open as in $\rm NGC\,2535$. Moreover, Gal2int is not barred but this is due to the mild resolution of our simulation (see discussion on the effect of resolution in Section~4.3.6 below).

\subsubsection{Kinematics and metal abundances}
The best match configuration for the gaseous component is presented in Fig.~\ref{fig:simugas_figure}. The upper panel shows surface density maps of the simulated galaxies. The other panels show the velocity maps, oxygen abundance maps, iron abundance maps, respectively for the same galaxies, but only considering ionized gas. The ionized gas is confined in the tidal features described above at $t=0.39\,\rm Gyr$. The velocity maps are obtained by plotting $v_z$ of the gas particles to which we add the average ($4\,121\,\rm km\,s^{-1}$) of the systemic velocities found for $\rm NGC\,2535$ and $\rm NGC\,2536$. Before the first pericentre passage, the lines of nodes of both Gal1int and Gal2int lie almost in the same direction. After the first pericentre passage, the gaseous disk of Gal2int experiences multiple deformations and is shattered, losing big chunks in its periphery which latter re-accrete making it difficult to analyze its rotation even though the central part of Gal2int in the velocity map in Fig.~\ref{fig:simugas_figure} seems to indicate normal rotation. In the metallicity maps, the gas re-accreting on Gal2int have a slightly lower metallicity than the central part which exhibits an almost uniform value caused by mixing. The gaseous debris do not accumulate at the tip of any tidal tail to form massive TDGs. Nevertheless, gaseous condensations form after the third pericentre passage. Corresponding overdensities in the stellar component can be seen in the snapshots of Fig.~\ref{fig:simustars_figure} at $t=0.65\,\rm Gyr$ near positions ($-1,4.5$), ($2.5,2.5$), and ($16,4.5$),
and $t=0.85\,\rm Gyr$ near positions ($-8,-1$), ($0,-3.5$), and ($6.5;-13.5$).
These objects survive until the end of the run. For Gal1int, the differential rotation is still distinguishable up until the second pericentre passage. The line of nodes lies almost in the direction joining the centres of Gal1int and Gal2int at the best match configuration. This explains the offset between the kinematic and photometric P.As in $\rm NGC\,2535$. After the second pericentre passage, the difference between velocities of the receding and approaching sides diminishes considerably, leaving at the end of the run a slow rotating remanent. The simulation reproduces satisfactorily the velocity and O/H distributions of $\rm NGC\,2535$ when comparing the velocity maps and O/H map of Gal1int in Fig.~\ref{fig:simugas_figure} to those of $\rm NGC\,2535$ presented in Fig.~\ref{fig:velocitymap_figure} and Fig.~\ref{fig:abundance_figure}.

\subsubsection{Local properties}
 To investigate the local effects of the interaction, we conduct a radial profile analysis of the 2D velocity maps of Gal1int and Gal1iso at $t=0.39\,\rm Gyr$ (using DiskFit) and of their 2D O/H maps. The results are shown in the top and middle panels of Fig.~\ref{fig:VOHprof_figure}. For comparison purposes, we re-derive the radial velocity profile of $\rm NGC\,2535$ to include the region beyond the ocular by constraining all kinematic parameters to the values reported in Table~\ref{tab:table2}. It is interesting to note that both Gal1int and $\rm NGC\,2535$ exhibit a steep decrease beyond $\sim25^{\prime\prime}$ whereas the velocity profile of Gal1iso reaches a plateau. The decrease is not continuous but rather reflects the constant velocities over long patches in the stretched tidal tails caused by streaming motions. The same effect explains the constant values for the metallicity over long patches of the tidal tails as well. Also of interest, there is a noticeable difference between the velocity profiles of Gal1int and Gal1iso in the inner $\sim15^{\prime\prime}$ where the amplitude is bigger in Gal1int. 
 Earlier simulations have revealed that ocular wave in interacting galaxies results from the inflow and speed up of material in the two quadrants that lose angular momentum as a result of a quasi-impulsive encounter at closest approach  \citep[][]{Curtis1999,Elmegreen1995,Elmegreen1991}. This inflow accounts for both the density increase and the centrifugal speed-up in Gal1int.
 The velocity profiles of Gal1int and $\rm NGC\,2535$ are very similar in shapes although we point out a slight difference in radial velocity values in the inner and outer parts of the ionized gas disks. From Fig.~\ref{fig:VOHprof_figure}, Gal1int and $\rm NGC\,2535$ have similar oxygen abundance profiles as well, both exhibiting a break beyond the radius of $\sim25^{\prime\prime}$ albeit with a big scatter in $\rm NGC\,2535$ values whereas the profile in Gal1iso could be fitted with a single slope. Inside this radius, O/H values of Gal1int are systematically lower than the values of Gal1iso although the difference is very little. Hence, the inner slopes of O/H vs. radius for Gal1int and Gal1iso would be comparable. Data points for the O/H profiles in the simulated galaxies are obtained by averaging over concentric annuli (centre, P.A. and ellipticity derived from the kinematic analysis) and error bars correspond to the standard deviation at a given radius. The data points from the observations are measured abundances of the HII region complexes detected in $\rm NGC\,2535$ and their corresponding uncertainties. O/H values over the entire disk of Gal2int are very close, predicting a shallow abundance profile, although the complexity and mismatch of the velocity map to that of $\rm NGC\,2536$ does not allow us to carry out the same radial analysis on Gal2int. We also investigated the distribution of the iron abundance in the simulated galaxies. The corresponding profiles, shown in the bottom panel of Fig.\ref{fig:VOHprof_figure}, are also obtained by averaging [Fe/H] values of Gal1int and Gal1iso over concentric annuli. [Fe/H] distribution could be fitted with a single slope for Gal1iso and two slopes for Gal1int. For the interacting galaxy, the outer slope, beyond the break radius, is still shallow. Inside the break radius, [Fe/H] values for Gal1int are still systematically lower than the values for Gal1iso but the difference is more noticeable compared to the O/H values. 

\subsubsection{Global properties}
Fig.~\ref{fig:globalevolution_figure} shows the time evolution of the global SFR, stellar and gaseous masses, iron and oxygen abundances. Before the first pericentre passage, the total SFR in the interacting system is almost constant at around $7.5\,M_\odot\,\rm yr^{-1}$ and is comparable to the SFR in Gal1iso. SFR in Gal2iso is negligible. The global SFR in isolated galaxies Gal1iso and Gal2iso stays almost constant throughout the run. After the first pericentre passage ($t=0.25\,\rm Gyr$), the global SFR in Gal1int+Gal2int increases linearly to reach $12.3\,M_\odot\,\rm yr^{-1}$ at $t=0.39\,\rm Gyr$, matching the value derived form observations. At this moment, the total stellar mass is $5.73\times10^{10}M_\odot$ and total gas mass is $3.22\times10^{10}\,M_\odot$, of which $2.3\times10^{10}M_\odot$ is hydrogen, matching the estimated masses from observations. Right before the second pericentre passage, at $t=0.43\,\rm Gyr$, it increases significantly to be $24\,M_\odot\,\rm yr^{-1}$ at the second pericentre passage, at $t=0.49\,\rm Gyr$. The global SFR is $33\,M_\odot\,\rm yr^{-1}$ by the time Gal1int and Gal2int are at their third closest separation at $t=0.63\,\rm Gyr$. Past the third pericentre passage, the global SFR increases drastically to reach its peak at $74\,M_\odot\,\rm yr^{-1}$ (ten times the corresponding value in Gal1int at the same time) in only 0.04 Gyr. After this, it decreases linearly. When the galaxies have coalesced, at $t=0.83\,\rm Gyr$, the value of the SFR of the system is comparable to that at the second pericentre passage, three times lower the value at its peak. The evolution of the stellar/gaseous masses in Gal1int+Gal2int is related to the SFR. The slope of the M$_\star$/M$_\mathrm{g}$ vs. time increase/decrease at $t=0.25$, 0.43 and $0.63\,\rm Gyr$. After $t=0.67\,\rm Gyr$, the changes in slopes are moderate even if the gaseous mass is 70 per cent its initial value : the available gas is heated and is not used to form new stars. The masses in the isolated galaxies change very little comparatively throughout the run as the corresponding SFRs are low. The bottom panels of Fig.~\ref{fig:globalevolution_figure} show the temporal evolution of O/H and Fe/H averaged over the entire systems (Gal1int+Gal2int, Gal1iso, and Gal2iso). The evolution of O/H is closely related to the evolution of the global SFR. At the beginning, O/H in Gal1int+Gal2int is similar to O/H in Gal1iso. O/H rises at rates that change almost at the times, as discussed above, where the global SFR accelerates but with a small delay. The evolution of Fe/H tends to be more steady, even in the isolated galaxies. These differences can be explained by the different lifetimes of Type II and Type Ia SNe progenitors. Since oxygen is mainly produced by Type II SNe, their very short lifetimes progenitors make the histories of SF and oxygen enrichment correlated. Type Ia SNe, which mainly contribute to the production of iron, have very long lifetimes progenitors. Then, a significant fraction of iron is produced long after the SFR has reached its peak, hence the notable difference between O/H and Fe/H histories.

\begin{figure}
    \includegraphics[scale=0.55,left]{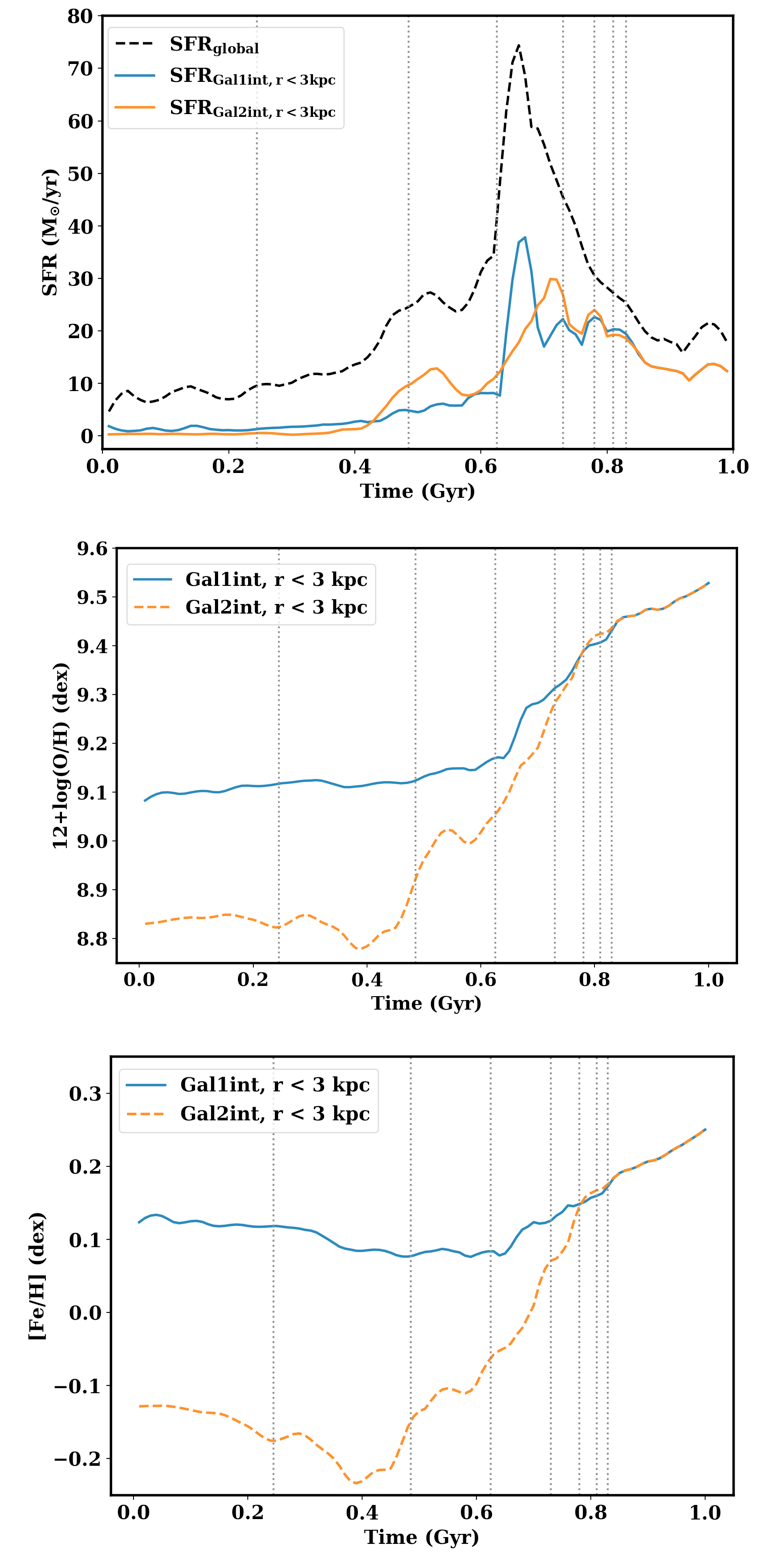}
    \caption{Temporal evolution of the central SFR (upper panel), iron and oxygen abundances (middle and lower panel respectively) of Gal1int and Gal2int.}
    \label{fig:centralevol_figure}
\end{figure}

\begin{figure*}
    \includegraphics[scale=0.40]{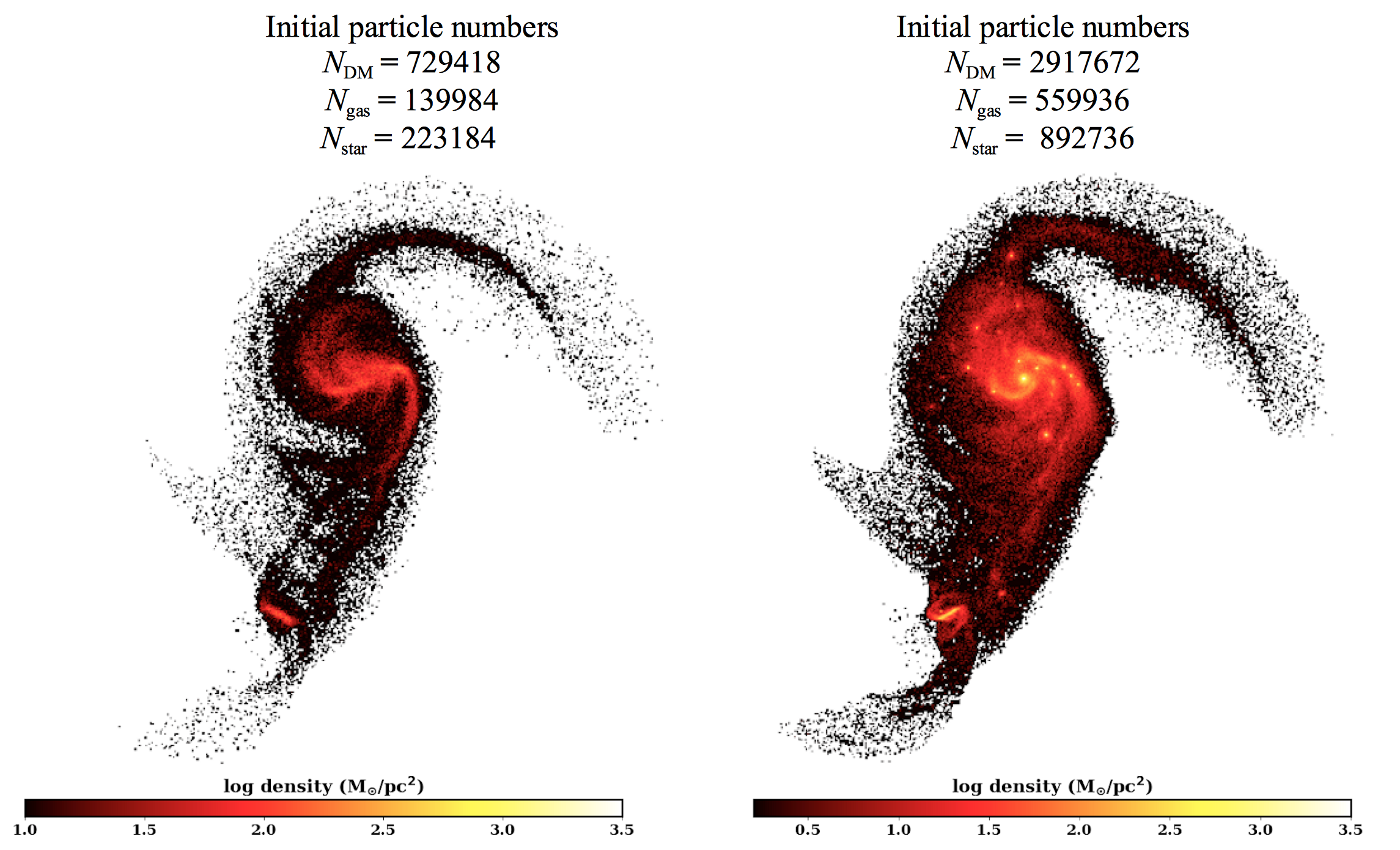}
    \caption{Stellar component snapshots at $t=0.39\,\rm Gyr$ in the $xy$ plane of the simulation presented in this work (left) and a higher resolution simulation (right). Top labels refer to the initial numbers of particles in the simulations.}
    \label{fig:simresolution_figure}
\end{figure*}

\subsubsection{Inside the central regions}
We also computed the central SFRs and metallicities to see how the evolution of these quantities compares to the global ones. Fig.~\ref{fig:centralevol_figure} shows the history of the SFR, O/H and Fe/H averaged over a spherical region of radius 3 kpc centered on the centres of mass of Gal1int and Gal2int. The central SFRs increase significantly at the third pericentre passage but Gal2int exhibits a burst of central SF at the second pericentre passage. The central SFRs in Gal1int and Gal2int after the third pericentre passage account for more than half of the global SFR. We note that the central SFR in Gal1int is more sensitive to the separation between the galaxies after the third pericentre passage as it exhibits enhancements when the galaxies are at their closest passages. The central chemical evolution differs from the global one. Initially, there is a drop in O/H and Fe/H caused by low metallicity gas channeled to the centre. The effect is more noticeable for Gal2int since the companion is more affected by the tidal forces as discussed above. The production of oxygen and iron (but to a lesser extent) by Type II SNe compensates this initial drop. The metals enrichment ends up surpassing the increase in hydrogen in the centre and drives up the rise in the history of Fe/H and O/H. This rise is correlated to the bursts of SF in Gal2int and Gal1int at the second and third pericentre passages respectively. There is a change of slope in O/H at $t=0.78\,\rm Gyr$ due to the decline of central SFRs at the same time. Since iron is produced mainly by Type Ia SNe with long lifetimes progenitors, the change of slope does not happen in the history of Fe/H. 

The limited resolution of the algorithm might be responsible for the high central abundances that are reached by the end of the simulation. The spike in SFR after the third pericentre passage should lead to significant feedback, that would regulate star formation. However, the decrease in central SFR after the spike is quite slow, suggesting that, once the central gas density reaches a certain value, the simulation underestimate the strength of feedback at the current resolution. We note that this occurs long after the present time $t=0.39\,\rm Gyr$. Hence, this does not impact our comparison between the simulation and the observations. 

\subsubsection{Effect of the resolution of the numerical simulation}

In the previous sections, we presented a suite of three numerical simulations: two of isolated galaxies, and one of interacting galaxies, with the goal of reproducing the observations of $\rm NGC\,2535$ and $\rm NGC\,2536$.
To test the robustness of the comparison, we started a fourth simulation of interacting galaxies, using the same initial conditions but four times as many gas and dark matter particles. The purpose of this high-resolution run is to allow a comparison with the low-resolution run and the observations at the present. We are particularly interested in small morphological features that are present in the observations, but not reproduced in the low-resolution simulation. Since the computational cost was getting excessive, we did not push this new simulation beyond $t=0.39\,\rm Gyr$.

Fig.~\ref{fig:simresolution_figure} shows snapshots at $t=0.39\,\rm Gyr$ of the stellar component in the plane of the sky, for the original simulation and the higher-resolution one.
The morphological structures induced by the interaction in the primary galaxy stand out more clearly in the higher resolution simulation and the satellite galaxy displays a bar and spiral arms that are more tightly wound. Furthermore, the thin dense region in the extended tidal arm on the anti-companion side is less curved and appears broken. The higher number of gas particles allows a better sampling of higher gas density regions leading to more stars being formed.
This explains the condensations or clumps in the stellar components, visible in the tidal features at high resolution (yellow spots in
Fig.~\ref{fig:simresolution_figure}). 

In this first high-resolution simulation, the SFR is higher than in the low-resolution one, because we did not adjust the value of the parameter $C*$ in equation~(\ref{eq:SF}) to account for the fact that, at higher resolution, the algorithm can resolve regions with higher gas densities. We defer this to future work, as our main objective here was to asses the effect of resolution on the morphology of the system.

\section{Summary and Conclusions}
\label{sec:section5}
 We presented kinematics and metallicity analysis of the interacting pair of star-forming galaxies Arp~82 using SITELLE data. We also investigated the effects of the interaction using a numerical model that reproduces some of the properties of the interacting galaxies $\rm NGC\,2535$ and $\rm NGC\,2536$. The kinematics and photometry analysis of $\rm NGC\,2536$ reveal that the satellite galaxy is barred. The observed velocity map of $\rm NGC\,2535$ is axisymmetric inside the ocular ring with a small bump in the inner $10^{\prime\prime}$ ($3\,\rm kpc$). The numerical model reproduces the shape of this velocity profile. Comparison to the kinematics of the simulated primary galaxy evolved in isolation shows that the bump is induced by the interaction. An increase in central densities during the interaction drives the rise in central velocities. Beyond the eye-shaped structure of $\rm NGC\,2535$, at $r>25^{\prime\prime}$ ($7\,\rm kpc$), we find strong streaming motions around the ocular and along the tidal tails with constant velocities over long patches in the velocity map and a corresponding strong decrease in the velocity profile of the receding side at this radius. Peculiar velocities or velocity kinks in the H$\upalpha$ velocity map of $\rm NGC\,2535$ bring out an other structure with an elliptical shape. This elliptical arc is also detected in SITELLE deep images. The numerical model reproduces the observed ocular, the spurs around the ocular, the elliptical arc, and tidal tails and shows that these features are induced by the interaction. It also reproduces the intrinsic ellipticity of the stellar disk of $\rm NGC\,2535$ induced by the interaction at the first close encounter in a direction almost perpendicular to the kinematic P.A. In fact, these features are produced after the galaxies have experienced a first collision, half way in their race to a second collision. Tidal forces at the first collision result in most of the outer gas of the satellite galaxy being pulled off, and the inner gas disk experiencing several distortions. The subsequent mixing leaves the residual inner disk with an almost uniform value of metallicity with some of the removed gas, with low metallicity, falling back on it. This could explain the very shallow oxygen abundance radial profile observed for $\rm NGC\,2536$. For the primary galaxy, at the best match configuration, only the outer disk is affected, with the tidal tails showing constant metallicity over long patches. This result in a break in the metallicity profile, similar to the one observed for $\rm NGC\,2535$.
 
The numerical model also reproduces the observed global SFR of Arp 82 and its total stellar/gas mass. The global SFR is about one and a half times that of the primary galaxy evolved in isolation. The corresponding central SFRs are not important, meaning that the induced SF is spatially extended. The central oxygen abundance of the satellite galaxy is lower compared to its initial value and its O/H radial profile is flattened by the interaction. However, for the primary galaxy, the central oxygen abundance does not change much from its initial value and its O/H radial profile, inside the ocular, is comparable to that of normal isolated star forming galaxies. The global SFR in the simulation reaches its peak when the two galaxies are still separated. At its peak, the global SFR is about ten times its value in the primary galaxy evolved in isolation, consistent with what is measured for bright local starbursts. When the galaxies have merged, the global SFR is three times lower compared to its peak value. The results presented in this work show that the changes induced by the interaction are a function of the interaction stage and galaxies properties. Early statistical studies investigated properties of interacting galaxies based on pair separations and relative velocity. Since the projected distances between pairs of galaxies does not really represent their physical separations and is not correlated to the interaction stage, detailed studies of specific nearby systems are crucial for a better characterization of the properties of interacting galaxies.  

\section*{Acknowledgements}
Based  on  observations  obtained  with  SITELLE,  a  joint project  of  Universit\'e  Laval,  ABB,  Universit\'e  de  Montr\'eal and  the  Canada-France-Hawaii  Telescope  (CFHT), which is  operated  by  the  National  Research  Council of Canada,  the  Institut  National  des  Sciences  de  l'Univers  of the  Centre  National  de  la  Recherche  Scientifique of France and the University of Hawaii. The authors wish to recognize and acknowledge the very significant cultural role that the summit of Mauna Kea has always had within
the indigenous Hawaiian community. We are most grateful to have
the opportunity to conduct observations from this mountain.
LD and HM are grateful to the Natural Sciences and Engineering Research Council of
Canada, the Fonds de Recherche du Qu\'ebec, 
and the Canada Foundation for Innovation for financial support. JIP and JVM acknowledge finantial support from projects Estallidos6 AYA2016-79724-C4 (Spanish Ministerio de Economia y Competitividad), Estallidos7 PID2019-107408GB-C44 (Spanish Ministerio de Ciencia e Innovacion), grant P18-FR-2664 (Junta de Andalucía), and grant SEV-2017-0709 “Centro de Excelencia Severo Ochoa Program” (Spanish Science Ministry).
We thank the referee, Dr. Curtis Struck, for a thorough review of this paper and valuable suggestions.

\section*{Data Availability Statement}
The data underlying this article will be shared on reasonable request to the authors.



\bibliographystyle{mnras}
\bibliography{references}

\newcommand{\noop}[1]{}
\begin{thebibliography}{}
\makeatletter
\relax
\def\mn@urlcharsother{\let\do\@makeother \do\$\do\&\do\#\do\^\do\_\do\%\do\~}
\def\mn@doi{\begingroup\mn@urlcharsother \@ifnextchar [ {\mn@doi@}
  {\mn@doi@[]}}
\def\mn@doi@[#1]#2{\def\@tempa{#1}\ifx\@tempa\@empty \href
  {http://dx.doi.org/#2} {doi:#2}\else \href {http://dx.doi.org/#2} {#1}\fi
  \endgroup}
\def\mn@eprint#1#2{\mn@eprint@#1:#2::\@nil}
\def\mn@eprint@arXiv#1{\href {http://arxiv.org/abs/#1} {{\tt arXiv:#1}}}
\def\mn@eprint@dblp#1{\href {http://dblp.uni-trier.de/rec/bibtex/#1.xml}
  {dblp:#1}}
\def\mn@eprint@#1:#2:#3:#4\@nil{\def\@tempa {#1}\def\@tempb {#2}\def\@tempc
  {#3}\ifx \@tempc \@empty \let \@tempc \@tempb \let \@tempb \@tempa \fi \ifx
  \@tempb \@empty \def\@tempb {arXiv}\fi \@ifundefined
  {mn@eprint@\@tempb}{\@tempb:\@tempc}{\expandafter \expandafter \csname
  mn@eprint@\@tempb\endcsname \expandafter{\@tempc}}}

\bibitem[\protect\citeauthoryear{{Amram}, {Marcelin}, {Boulesteix}  \& {Le
  Coarer}}{{Amram} et~al.}{1989}]{Amram1989}
{Amram} P.,  {Marcelin} M.,  {Boulesteix} J.,   {Le Coarer} E.,  1989, \aaps,
  \href {https://ui.adsabs.harvard.edu/abs/1989A&AS...81...59A} {81, 59}

\bibitem[\protect\citeauthoryear{{Baldwin}, {Phillips}  \&
  {Terlevich}}{{Baldwin} et~al.}{1981}]{BPT1981}
{Baldwin} J.~A.,  {Phillips} M.~M.,   {Terlevich} R.,  1981, \mn@doi [\pasp]
  {10.1086/130766}, \href
  {https://ui.adsabs.harvard.edu/abs/1981PASP...93....5B} {93, 5}

\bibitem[\protect\citeauthoryear{{Barrera-Ballesteros}
  et~al.,}{{Barrera-Ballesteros} et~al.}{2015}]{Barrera-Ballesteros2015}
{Barrera-Ballesteros} J.~K.,  et~al., 2015, \mn@doi [\aap]
  {10.1051/0004-6361/201425397}, \href
  {https://ui.adsabs.harvard.edu/abs/2015A&A...579A..45B} {579, A45}

\bibitem[\protect\citeauthoryear{{Blumenthal} \& {Barnes}}{{Blumenthal} \&
  {Barnes}}{2018}]{Blumenthal2018}
{Blumenthal} K.~A.,  {Barnes} J.~E.,  2018, \mn@doi [\mnras]
  {10.1093/mnras/sty1605}, \href
  {https://ui.adsabs.harvard.edu/abs/2018MNRAS.479.3952B} {479, 3952}

\bibitem[\protect\citeauthoryear{{Bournaud}, {Duc}, {Amram}, {Combes}  \&
  {Gach}}{{Bournaud} et~al.}{2004}]{Bournaud2004}
{Bournaud} F.,  {Duc} P.~A.,  {Amram} P.,  {Combes} F.,   {Gach} J.~L.,  2004,
  \mn@doi [\aap] {10.1051/0004-6361:20040394}, \href
  {https://ui.adsabs.harvard.edu/abs/2004A&A...425..813B} {425, 813}

\bibitem[\protect\citeauthoryear{{Bresolin}}{{Bresolin}}{2007}]{Bresolin2007}
{Bresolin} F.,  2007, \mn@doi [\apj] {10.1086/510380}, \href
  {https://ui.adsabs.harvard.edu/abs/2007ApJ...656..186B} {656, 186}

\bibitem[\protect\citeauthoryear{{Buat} \& {Xu}}{{Buat} \&
  {Xu}}{1996}]{Buat1996}
{Buat} V.,  {Xu} C.,  1996, \aap, \href
  {https://ui.adsabs.harvard.edu/abs/1996A&A...306...61B} {306, 61}

\bibitem[\protect\citeauthoryear{{Calzetti}}{{Calzetti}}{2001}]{Calzetti2001}
{Calzetti} D.,  2001, \mn@doi [\pasp] {10.1086/324269}, \href
  {https://ui.adsabs.harvard.edu/abs/2001PASP..113.1449C} {113, 1449}

\bibitem[\protect\citeauthoryear{{Cardelli}, {Clayton}  \& {Mathis}}{{Cardelli}
  et~al.}{1989}]{Cardelli1989}
{Cardelli} J.~A.,  {Clayton} G.~C.,   {Mathis} J.~S.,  1989, \mn@doi [\apj]
  {10.1086/167900}, \href
  {https://ui.adsabs.harvard.edu/abs/1989ApJ...345..245C} {345, 245}

\bibitem[\protect\citeauthoryear{{Courteau}}{{Courteau}}{1997}]{Courteau1997}
{Courteau} S.,  1997, \mn@doi [\aj] {10.1086/118656}, \href
  {https://ui.adsabs.harvard.edu/abs/1997AJ....114.2402C} {114, 2402}

\bibitem[\protect\citeauthoryear{{Di Matteo}, {Combes}, {Melchior}  \&
  {Semelin}}{{Di Matteo} et~al.}{2007}]{DiMatteo2007}
{Di Matteo} P.,  {Combes} F.,  {Melchior} A.~L.,   {Semelin} B.,  2007, \mn@doi
  [\aap] {10.1051/0004-6361:20066959}, \href
  {https://ui.adsabs.harvard.edu/abs/2007A&A...468...61D} {468, 61}

\bibitem[\protect\citeauthoryear{{Drew}, {Casey}, {Burnham}, {Hung}, {Kassin},
  {Simons}  \& {Zavala}}{{Drew} et~al.}{2018}]{Drew2018}
{Drew} P.~M.,  {Casey} C.~M.,  {Burnham} A.~D.,  {Hung} C.-L.,  {Kassin} S.~A.,
   {Simons} R.~C.,   {Zavala} J.~A.,  2018, \mn@doi [\apj]
  {10.3847/1538-4357/aaedbf}, \href
  {https://ui.adsabs.harvard.edu/abs/2018ApJ...869...58D} {869, 58}

\bibitem[\protect\citeauthoryear{{Drissen}, {Rousseau-Nepton}, {Lavoie},
  {Robert}, {Martin}, {Martin}, {Mandar}  \& {Grandmont}}{{Drissen}
  et~al.}{2014}]{Drissen2014}
{Drissen} L.,  {Rousseau-Nepton} L.,  {Lavoie} S.,  {Robert} C.,  {Martin} T.,
  {Martin} P.,  {Mandar} J.,   {Grandmont} F.,  2014, \mn@doi [Advances in
  Astronomy] {10.1155/2014/293856}, \href
  {https://ui.adsabs.harvard.edu/abs/2014AdAst2014E...9D} {2014, 293856}

\bibitem[\protect\citeauthoryear{{Drissen} et~al.,}{{Drissen}
  et~al.}{2019}]{Drissen2019}
{Drissen} L.,  et~al., 2019, \mn@doi [\mnras] {10.1093/mnras/stz627}, \href
  {https://ui.adsabs.harvard.edu/abs/2019MNRAS.485.3930D} {485, 3930}

\bibitem[\protect\citeauthoryear{{Duc}}{{Duc}}{2012}]{Duc2012}
{Duc} P.-A.,  2012, \mn@doi [Astrophysics and Space Science Proceedings]
  {10.1007/978-3-642-22018-0_37}, \href
  {https://ui.adsabs.harvard.edu/abs/2012ASSP...28..305D} {28, 305}

\bibitem[\protect\citeauthoryear{{Elbaz} et~al.,}{{Elbaz}
  et~al.}{2007}]{Elbaz2007}
{Elbaz} D.,  et~al., 2007, \mn@doi [\aap] {10.1051/0004-6361:20077525}, \href
  {https://ui.adsabs.harvard.edu/abs/2007A&A...468...33E} {468, 33}

\bibitem[\protect\citeauthoryear{{Ellison}, {Mendel}, {Patton}  \&
  {Scudder}}{{Ellison} et~al.}{2013}]{Ellison2013}
{Ellison} S.~L.,  {Mendel} J.~T.,  {Patton} D.~R.,   {Scudder} J.~M.,  2013,
  \mn@doi [\mnras] {10.1093/mnras/stt1562}, \href
  {https://ui.adsabs.harvard.edu/abs/2013MNRAS.435.3627E} {435, 3627}

\bibitem[\protect\citeauthoryear{{Elmegreen}, {Sundin}, {Elmegreen}  \&
  {Sundelius}}{{Elmegreen} et~al.}{1991}]{Elmegreen1991}
{Elmegreen} D.~M.,  {Sundin} M.,  {Elmegreen} B.,   {Sundelius} B.,  1991,
  \aap, \href {https://ui.adsabs.harvard.edu/abs/1991A&A...244...52E} {244, 52}

\bibitem[\protect\citeauthoryear{{Elmegreen}, {Kaufman}, {Brinks}, {Elmegreen}
  \& {Sundin}}{{Elmegreen} et~al.}{1995}]{Elmegreen1995}
{Elmegreen} D.~M.,  {Kaufman} M.,  {Brinks} E.,  {Elmegreen} B.~G.,   {Sundin}
  M.,  1995, \mn@doi [\apj] {10.1086/176374}, \href
  {https://ui.adsabs.harvard.edu/abs/1995ApJ...453..100E} {453, 100}

\bibitem[\protect\citeauthoryear{{Hancock}, {Smith}, {Struck}, {Giroux},
  {Appleton}, {Charmand aris}  \& {Reach}}{{Hancock}
  et~al.}{2007}]{Hancock2007}
{Hancock} M.,  {Smith} B.~J.,  {Struck} C.,  {Giroux} M.~L.,  {Appleton} P.~N.,
   {Charmand aris} V.,   {Reach} W.~T.,  2007, \mn@doi [\aj] {10.1086/510241},
  \href {https://ui.adsabs.harvard.edu/abs/2007AJ....133..676H} {133, 676}

\bibitem[\protect\citeauthoryear{{Holincheck} et~al.,}{{Holincheck}
  et~al.}{2016}]{Holincheck2016}
{Holincheck} A.~J.,  et~al., 2016, \mn@doi [\mnras] {10.1093/mnras/stw649},
  \href {https://ui.adsabs.harvard.edu/abs/2016MNRAS.459..720H} {459, 720}

\bibitem[\protect\citeauthoryear{{Howard}, {Keel}, {Byrd}  \&
  {Burkey}}{{Howard} et~al.}{1993}]{Howard1993}
{Howard} S.,  {Keel} W.~C.,  {Byrd} G.,   {Burkey} J.,  1993, \mn@doi [\apj]
  {10.1086/173329}, \href
  {https://ui.adsabs.harvard.edu/abs/1993ApJ...417..502H} {417, 502}

\bibitem[\protect\citeauthoryear{{Kauffmann} et~al.,}{{Kauffmann}
  et~al.}{2003}]{Kauffmann2003}
{Kauffmann} G.,  et~al., 2003, \mn@doi [\mnras]
  {10.1111/j.1365-2966.2003.07154.x}, \href
  {https://ui.adsabs.harvard.edu/abs/2003MNRAS.346.1055K} {346, 1055}

\bibitem[\protect\citeauthoryear{{Kaufman}, {Brinks}, {Elmegreen}, {Thomasson},
  {Elmegreen}, {Struck}  \& {Klaric}}{{Kaufman} et~al.}{1997}]{Kaufman1997}
{Kaufman} M.,  {Brinks} E.,  {Elmegreen} D.~M.,  {Thomasson} M.,  {Elmegreen}
  B.~G.,  {Struck} C.,   {Klaric} M.,  1997, \mn@doi [\aj] {10.1086/118651},
  \href {https://ui.adsabs.harvard.edu/abs/1997AJ....114.2323K} {114, 2323}

\bibitem[\protect\citeauthoryear{{Kawata} \& {Gibson}}{{Kawata} \&
  {Gibson}}{2003}]{Kawata2003}
{Kawata} D.,  {Gibson} B.~K.,  2003, \mn@doi [\mnras]
  {10.1046/j.1365-8711.2003.06356.x}, \href
  {https://ui.adsabs.harvard.edu/abs/2003MNRAS.340..908K} {340, 908}

\bibitem[\protect\citeauthoryear{{Kawata}, {Okamoto}, {Gibson}, {Barnes}  \&
  {Cen}}{{Kawata} et~al.}{2013}]{Kawata2013}
{Kawata} D.,  {Okamoto} T.,  {Gibson} B.~K.,  {Barnes} D.~J.,   {Cen} R.,
  2013, \mn@doi [\mnras] {10.1093/mnras/sts161}, \href
  {https://ui.adsabs.harvard.edu/abs/2013MNRAS.428.1968K} {428, 1968}

\bibitem[\protect\citeauthoryear{{Kennicutt}}{{Kennicutt}}{1998}]{Kennicutt1998ARA&A}
{Kennicutt} Robert~C. J.,  1998, \mn@doi [\araa]
  {10.1146/annurev.astro.36.1.189}, \href
  {https://ui.adsabs.harvard.edu/abs/1998ARA&A..36..189K} {36, 189}

\bibitem[\protect\citeauthoryear{{Kennicutt}, {Keel}, {van der Hulst}, {Hummel}
   \& {Roettiger}}{{Kennicutt} et~al.}{1987}]{Kennicutt1987}
{Kennicutt} Robert~C. J.,  {Keel} W.~C.,  {van der Hulst} J.~M.,  {Hummel} E.,
   {Roettiger} K.~A.,  1987, \mn@doi [\aj] {10.1086/114384}, \href
  {https://ui.adsabs.harvard.edu/abs/1987AJ.....93.1011K} {93, 1011}

\bibitem[\protect\citeauthoryear{{Kewley} \& {Dopita}}{{Kewley} \&
  {Dopita}}{2002}]{Kewley2002}
{Kewley} L.~J.,  {Dopita} M.~A.,  2002, \mn@doi [\apjs] {10.1086/341326}, \href
  {https://ui.adsabs.harvard.edu/abs/2002ApJS..142...35K} {142, 35}

\bibitem[\protect\citeauthoryear{{Kewley} \& {Ellison}}{{Kewley} \&
  {Ellison}}{2008}]{KewleyEllison2008}
{Kewley} L.~J.,  {Ellison} S.~L.,  2008, \mn@doi [\apj] {10.1086/587500}, \href
  {https://ui.adsabs.harvard.edu/abs/2008ApJ...681.1183K} {681, 1183}

\bibitem[\protect\citeauthoryear{{Kewley}, {Dopita}, {Sutherland}, {Heisler}
  \& {Trevena}}{{Kewley} et~al.}{2001}]{Kewley2001}
{Kewley} L.~J.,  {Dopita} M.~A.,  {Sutherland} R.~S.,  {Heisler} C.~A.,
  {Trevena} J.,  2001, \mn@doi [\apj] {10.1086/321545}, \href
  {https://ui.adsabs.harvard.edu/abs/2001ApJ...556..121K} {556, 121}

\bibitem[\protect\citeauthoryear{{Kewley}, {Geller}  \& {Barton}}{{Kewley}
  et~al.}{2006}]{Kewley2006}
{Kewley} L.~J.,  {Geller} M.~J.,   {Barton} E.~J.,  2006, \mn@doi [\aj]
  {10.1086/500295}, \href
  {https://ui.adsabs.harvard.edu/abs/2006AJ....131.2004K} {131, 2004}

\bibitem[\protect\citeauthoryear{{Kewley}, {Rupke}, {Zahid}, {Geller}  \&
  {Barton}}{{Kewley} et~al.}{2010}]{Kewley2010}
{Kewley} L.~J.,  {Rupke} D.,  {Zahid} H.~J.,  {Geller} M.~J.,   {Barton} E.~J.,
   2010, \mn@doi [\apjl] {10.1088/2041-8205/721/1/L48}, \href
  {https://ui.adsabs.harvard.edu/abs/2010ApJ...721L..48K} {721, L48}

\bibitem[\protect\citeauthoryear{{Knapen}, {Cisternas}  \&
  {Querejeta}}{{Knapen} et~al.}{2015}]{Knapen2015}
{Knapen} J.~H.,  {Cisternas} M.,   {Querejeta} M.,  2015, \mn@doi [\mnras]
  {10.1093/mnras/stv2135}, \href
  {https://ui.adsabs.harvard.edu/abs/2015MNRAS.454.1742K} {454, 1742}

\bibitem[\protect\citeauthoryear{{Kobulnicky} \& {Kewley}}{{Kobulnicky} \&
  {Kewley}}{2004}]{Kobulnicky2004}
{Kobulnicky} H.~A.,  {Kewley} L.~J.,  2004, \mn@doi [\apj] {10.1086/425299},
  \href {https://ui.adsabs.harvard.edu/abs/2004ApJ...617..240K} {617, 240}

\bibitem[\protect\citeauthoryear{{Kumari}, {Maiolino}, {Belfiore}  \&
  {Curti}}{{Kumari} et~al.}{2019}]{Kumari2019}
{Kumari} N.,  {Maiolino} R.,  {Belfiore} F.,   {Curti} M.,  2019, \mn@doi
  [\mnras] {10.1093/mnras/stz366}, \href
  {https://ui.adsabs.harvard.edu/abs/2019MNRAS.485..367K} {485, 367}

\bibitem[\protect\citeauthoryear{{Lee}, {Hwang}, {Lee}, {Kim}  \& {Lee}}{{Lee}
  et~al.}{2012}]{Lee2012}
{Lee} J.~C.,  {Hwang} H.~S.,  {Lee} M.~G.,  {Kim} M.,   {Lee} J.~H.,  2012,
  \mn@doi [\apj] {10.1088/0004-637X/756/1/95}, \href
  {https://ui.adsabs.harvard.edu/abs/2012ApJ...756...95L} {756, 95}

\bibitem[\protect\citeauthoryear{{Marino} et~al.,}{{Marino}
  et~al.}{2013}]{Marino2013}
{Marino} R.~A.,  et~al., 2013, \mn@doi [\aap] {10.1051/0004-6361/201321956},
  \href {https://ui.adsabs.harvard.edu/abs/2013A&A...559A.114M} {559, A114}

\bibitem[\protect\citeauthoryear{{Martin}, {Drissen}  \& {Joncas}}{{Martin}
  et~al.}{2015}]{Martin2015}
{Martin} T.,  {Drissen} L.,   {Joncas} G.,  2015, in {Taylor} A.~R.,
  {Rosolowsky} E.,  eds,  Astronomical Society of the Pacific Conference Series
  Vol. 495, Astronomical Data Analysis Software an Systems XXIV (ADASS XXIV).
  p.~327

\bibitem[\protect\citeauthoryear{{Martin}, {Prunet}  \& {Drissen}}{{Martin}
  et~al.}{2016}]{Martin2016}
{Martin} T.~B.,  {Prunet} S.,   {Drissen} L.,  2016, \mn@doi [\mnras]
  {10.1093/mnras/stw2315}, \href
  {https://ui.adsabs.harvard.edu/abs/2016MNRAS.463.4223M} {463, 4223}

\bibitem[\protect\citeauthoryear{{Martin}, {Drissen}  \& {Melchior}}{{Martin}
  et~al.}{2018}]{Martin2018}
{Martin} T.~B.,  {Drissen} L.,   {Melchior} A.-L.,  2018, \mn@doi [\mnras]
  {10.1093/mnras/stx2513}, \href
  {https://ui.adsabs.harvard.edu/abs/2018MNRAS.473.4130M} {473, 4130}

\bibitem[\protect\citeauthoryear{{Martin}, {Drissen}  \& {Prunet}}{{Martin}
  et~al.}{2021}]{Martin2021}
{Martin} T.,  {Drissen} L.,   {Prunet} S.,  2021, \mn@doi [\mnras]
  {10.1093/mnras/stab1656}, \href
  {https://ui.adsabs.harvard.edu/abs/2021MNRAS.505.5514M} {505, 5514}

\bibitem[\protect\citeauthoryear{{Mihos} \& {Hernquist}}{{Mihos} \&
  {Hernquist}}{1994}]{Mihos1994}
{Mihos} J.~C.,  {Hernquist} L.,  1994, \mn@doi [\apjl] {10.1086/187299}, \href
  {https://ui.adsabs.harvard.edu/abs/1994ApJ...425L..13M} {425, L13}

\bibitem[\protect\citeauthoryear{{Murphy} et~al.,}{{Murphy}
  et~al.}{2011}]{Murphy2011}
{Murphy} E.~J.,  et~al., 2011, \mn@doi [\apj] {10.1088/0004-637X/737/2/67},
  \href {https://ui.adsabs.harvard.edu/abs/2011ApJ...737...67M} {737, 67}

\bibitem[\protect\citeauthoryear{{Navarro}, {Frenk}  \& {White}}{{Navarro}
  et~al.}{1996}]{NFW1996}
{Navarro} J.~F.,  {Frenk} C.~S.,   {White} S. D.~M.,  1996, \mn@doi [\apj]
  {10.1086/177173}, \href
  {https://ui.adsabs.harvard.edu/abs/1996ApJ...462..563N} {462, 563}

\bibitem[\protect\citeauthoryear{{Osterbrock} \& {Ferland}}{{Osterbrock} \&
  {Ferland}}{2006}]{Osterbrock&Ferland2006}
{Osterbrock} D.~E.,  {Ferland} G.~J.,  2006, {Astrophysics of gaseous nebulae
  and active galactic nuclei}.
University Science Books

\bibitem[\protect\citeauthoryear{{P{\'e}rez-Montero}}{{P{\'e}rez-Montero}}{2014}]{PerezMontero2014}
{P{\'e}rez-Montero} E.,  2014, \mn@doi [\mnras] {10.1093/mnras/stu753}, \href
  {https://ui.adsabs.harvard.edu/abs/2014MNRAS.441.2663P} {441, 2663}

\bibitem[\protect\citeauthoryear{{P{\'e}rez-Montero}, {Garc{\'\i}a-Benito},
  {H{\"a}gele}  \& {D{\'\i}az}}{{P{\'e}rez-Montero}
  et~al.}{2010}]{PerezMontero2010}
{P{\'e}rez-Montero} E.,  {Garc{\'\i}a-Benito} R.,  {H{\"a}gele} G.~F.,
  {D{\'\i}az} {\'A}.~I.,  2010, \mn@doi [\mnras]
  {10.1111/j.1365-2966.2010.16421.x}, \href
  {https://ui.adsabs.harvard.edu/abs/2010MNRAS.404.2037P} {404, 2037}

\bibitem[\protect\citeauthoryear{{Persic} \& {Salucci}}{{Persic} \&
  {Salucci}}{1991}]{Persic1991}
{Persic} M.,  {Salucci} P.,  1991, \mn@doi [\apj] {10.1086/169670}, \href
  {https://ui.adsabs.harvard.edu/abs/1991ApJ...368...60P} {368, 60}

\bibitem[\protect\citeauthoryear{{Pilyugin} \& {Grebel}}{{Pilyugin} \&
  {Grebel}}{2016}]{Pilyugin2016}
{Pilyugin} L.~S.,  {Grebel} E.~K.,  2016, \mn@doi [\mnras]
  {10.1093/mnras/stw238}, \href
  {https://ui.adsabs.harvard.edu/abs/2016MNRAS.457.3678P} {457, 3678}

\bibitem[\protect\citeauthoryear{{Privon}, {Barnes}, {Evans}, {Hibbard}, {Yun},
  {Mazzarella}, {Armus}  \& {Surace}}{{Privon} et~al.}{2013}]{Privon2013}
{Privon} G.~C.,  {Barnes} J.~E.,  {Evans} A.~S.,  {Hibbard} J.~E.,  {Yun}
  M.~S.,  {Mazzarella} J.~M.,  {Armus} L.,   {Surace} J.,  2013, \mn@doi [\apj]
  {10.1088/0004-637X/771/2/120}, \href
  {https://ui.adsabs.harvard.edu/abs/2013ApJ...771..120P} {771, 120}

\bibitem[\protect\citeauthoryear{{Renaud}, {Bournaud}  \& {Duc}}{{Renaud}
  et~al.}{2015}]{Renaud2015}
{Renaud} F.,  {Bournaud} F.,   {Duc} P.-A.,  2015, \mn@doi [\mnras]
  {10.1093/mnras/stu2208}, \href
  {https://ui.adsabs.harvard.edu/abs/2015MNRAS.446.2038R} {446, 2038}

\bibitem[\protect\citeauthoryear{{Rupke}, {Kewley}  \& {Barnes}}{{Rupke}
  et~al.}{2010}]{Rupke2010}
{Rupke} D. S.~N.,  {Kewley} L.~J.,   {Barnes} J.~E.,  2010, \mn@doi [\apjl]
  {10.1088/2041-8205/710/2/L156}, \href
  {https://ui.adsabs.harvard.edu/abs/2010ApJ...710L.156R} {710, L156}

\bibitem[\protect\citeauthoryear{{Salpeter}}{{Salpeter}}{1955}]{Salpeter1955}
{Salpeter} E.~E.,  1955, \mn@doi [\apj] {10.1086/145971}, \href
  {https://ui.adsabs.harvard.edu/abs/1955ApJ...121..161S} {121, 161}

\bibitem[\protect\citeauthoryear{{S{\'a}nchez} et~al.,}{{S{\'a}nchez}
  et~al.}{2014}]{Sanchez2014}
{S{\'a}nchez} S.~F.,  et~al., 2014, \mn@doi [\aap]
  {10.1051/0004-6361/201322343}, \href
  {https://ui.adsabs.harvard.edu/abs/2014A&A...563A..49S} {563, A49}

\bibitem[\protect\citeauthoryear{{Sanders} \& {Mirabel}}{{Sanders} \&
  {Mirabel}}{1996}]{Sanders&Mirabel1996}
{Sanders} D.~B.,  {Mirabel} I.~F.,  1996, \mn@doi [\araa]
  {10.1146/annurev.astro.34.1.749}, \href
  {https://ui.adsabs.harvard.edu/abs/1996ARA&A..34..749S} {34, 749}

\bibitem[\protect\citeauthoryear{{Schawinski}, {Thomas}, {Sarzi}, {Maraston},
  {Kaviraj}, {Joo}, {Yi}  \& {Silk}}{{Schawinski}
  et~al.}{2007}]{Schawinski2007}
{Schawinski} K.,  {Thomas} D.,  {Sarzi} M.,  {Maraston} C.,  {Kaviraj} S.,
  {Joo} S.-J.,  {Yi} S.~K.,   {Silk} J.,  2007, \mn@doi [\mnras]
  {10.1111/j.1365-2966.2007.12487.x}, \href
  {https://ui.adsabs.harvard.edu/abs/2007MNRAS.382.1415S} {382, 1415}

\bibitem[\protect\citeauthoryear{{Sellwood} \& {S{\'a}nchez}}{{Sellwood} \&
  {S{\'a}nchez}}{2010}]{Sellwood2010}
{Sellwood} J.~A.,  {S{\'a}nchez} R.~Z.,  2010, \mn@doi [\mnras]
  {10.1111/j.1365-2966.2010.16430.x}, \href
  {https://ui.adsabs.harvard.edu/abs/2010MNRAS.404.1733S} {404, 1733}

\bibitem[\protect\citeauthoryear{{Sellwood} \& {Spekkens}}{{Sellwood} \&
  {Spekkens}}{2015}]{Sellwood2015}
{Sellwood} J.~A.,  {Spekkens} K.,  2015, arXiv e-prints, \href
  {https://ui.adsabs.harvard.edu/abs/2015arXiv150907120S} {p. arXiv:1509.07120}

\bibitem[\protect\citeauthoryear{{Spekkens} \& {Sellwood}}{{Spekkens} \&
  {Sellwood}}{2007}]{Spekkens2007}
{Spekkens} K.,  {Sellwood} J.~A.,  2007, \mn@doi [\apj] {10.1086/518471}, \href
  {https://ui.adsabs.harvard.edu/abs/2007ApJ...664..204S} {664, 204}

\bibitem[\protect\citeauthoryear{{Struck}}{{Struck}}{1999}]{Curtis1999}
{Struck} C.,  1999, \mn@doi [\physrep] {10.1016/S0370-1573(99)00030-7}, \href
  {https://ui.adsabs.harvard.edu/abs/1999PhR...321....1S} {321, 1}

\bibitem[\protect\citeauthoryear{{Thilker}, {Braun}  \& {Walterbos}}{{Thilker}
  et~al.}{2000}]{Thilker2000}
{Thilker} D.~A.,  {Braun} R.,   {Walterbos} R. A.~M.,  2000, \mn@doi [\aj]
  {10.1086/316852}, \href
  {https://ui.adsabs.harvard.edu/abs/2000AJ....120.3070T} {120, 3070}

\bibitem[\protect\citeauthoryear{{Thilker}, {Walterbos}, {Braun}  \&
  {Hoopes}}{{Thilker} et~al.}{2002}]{Thilker2002}
{Thilker} D.~A.,  {Walterbos} R. A.~M.,  {Braun} R.,   {Hoopes} C.~G.,  2002,
  \mn@doi [\aj] {10.1086/344303}, \href
  {https://ui.adsabs.harvard.edu/abs/2002AJ....124.3118T} {124, 3118}

\bibitem[\protect\citeauthoryear{{Toomre}}{{Toomre}}{1977}]{Toomre1977}
{Toomre} A.,  1977, in {Tinsley} B.~M.,  {Larson} R. B.~Gehret D.~C.,  eds,
  Evolution of Galaxies and Stellar Populations. p.~401

\bibitem[\protect\citeauthoryear{{Toomre} \& {Toomre}}{{Toomre} \&
  {Toomre}}{1972}]{Toomre&Toomre1972}
{Toomre} A.,  {Toomre} J.,  1972, \mn@doi [\apj] {10.1086/151823}, \href
  {https://ui.adsabs.harvard.edu/abs/1972ApJ...178..623T} {178, 623}

\bibitem[\protect\citeauthoryear{{Torres-Flores}, {Scarano}, {Mendes de
  Oliveira}, {de Mello}, {Amram}  \& {Plana}}{{Torres-Flores}
  et~al.}{2014}]{Torres-Flores2014}
{Torres-Flores} S.,  {Scarano} S.,  {Mendes de Oliveira} C.,  {de Mello} D.~F.,
   {Amram} P.,   {Plana} H.,  2014, \mn@doi [\mnras] {10.1093/mnras/stt2340},
  \href {https://ui.adsabs.harvard.edu/abs/2014MNRAS.438.1894T} {438, 1894}

\bibitem[\protect\citeauthoryear{{Weilbacher}, {Duc}  \& {Fritze-v.
  Alvensleben}}{{Weilbacher} et~al.}{2003}]{Weilbacher2003}
{Weilbacher} P.~M.,  {Duc} P.~A.,   {Fritze-v. Alvensleben} U.,  2003, \mn@doi
  [\aap] {10.1051/0004-6361:20021522}, \href
  {https://ui.adsabs.harvard.edu/abs/2003A&A...397..545W} {397, 545}

\bibitem[\protect\citeauthoryear{{Zaragoza-Cardiel}, {Smith}, {Rosado},
  {Beckman}, {Bitsakis}, {Camps-Fari{\~n}a}, {Font}  \&
  {Cox}}{{Zaragoza-Cardiel} et~al.}{2018}]{Zaragoza2018}
{Zaragoza-Cardiel} J.,  {Smith} B.~J.,  {Rosado} M.,  {Beckman} J.~E.,
  {Bitsakis} T.,  {Camps-Fari{\~n}a} A.,  {Font} J.,   {Cox} I.~S.,  2018,
  \mn@doi [\apjs] {10.3847/1538-4365/aaa255}, \href
  {https://ui.adsabs.harvard.edu/abs/2018ApJS..234...35Z} {234, 35}

\bibitem[\protect\citeauthoryear{{Zasov}, {Saburova}, {Egorov}  \&
  {Dodonov}}{{Zasov} et~al.}{2019}]{Zasov2019}
{Zasov} A.~V.,  {Saburova} A.~S.,  {Egorov} O.~V.,   {Dodonov} S.~N.,  2019,
  \mn@doi [\mnras] {10.1093/mnras/stz1025}, \href
  {https://ui.adsabs.harvard.edu/abs/2019MNRAS.486.2604Z} {486, 2604}

\bibitem[\protect\citeauthoryear{{Zibetti}, {Charlot}  \& {Rix}}{{Zibetti}
  et~al.}{2009}]{Zibetti2009}
{Zibetti} S.,  {Charlot} S.,   {Rix} H.-W.,  2009, \mn@doi [\mnras]
  {10.1111/j.1365-2966.2009.15528.x}, \href
  {https://ui.adsabs.harvard.edu/abs/2009MNRAS.400.1181Z} {400, 1181}

\makeatother
\end{thebibliography}

%%%%%%%%%%%%%%%%%%%%%%%%%%%%%%%%%%%%%%%%%%%%%%%%%%

%%%%%%%%%%%%%%%%% APPENDICES %%%%%%%%%%%%%%%%%%%%%

\appendix
\onecolumn
\section{Additional table and figure}
\label{sec:appendixA}
\begin{longtable}{cccccccccc}
    \caption{Positions and fluxes (uncorrected for reddening) of the HII region complexes detected in Arp 82. Fluxes are in units of $10^{-16}\,\mathrm{erg\,s^{-1}\,cm^{-2}}$.}\label{tab:tableA1}
    %\centering
     \\ \hline
    \small \makecell{Region \\ ID} & 
    \small \makecell{R.A} &
    \small \makecell{Dec} &
    \small \makecell{[OII]$\uplambda$3727} &
    \small \makecell{H$\upbeta$} &
    \small \makecell{[OIII]$\uplambda$5007} &
    \small \makecell{H$\upalpha$} &
    \small \makecell{[NII]$\uplambda$6583} &
    \small \makecell{[SII]$\uplambda$6716} &
    \small \makecell{[SII]$\uplambda$6731} \\
    \hline
    \endfirsthead
    \caption*{Table \ref{tab:tableA1}: Cont.} \\
    \hline
    \small \makecell{Region \\ ID} & 
    \small \makecell{R.A} &
    \small \makecell{Dec} &
    \small \makecell{[OII]$\uplambda$3727} &
    \small \makecell{H$\upbeta$} &
    \small \makecell{[OIII]$\uplambda$5007} &
    \small \makecell{H$\upalpha$} &
    \small \makecell{[NII]$\uplambda$6583} &
    \small \makecell{[SII]$\uplambda$6716} &
    \small \makecell{[SII]$\uplambda$6731} \\
    \hline
    \endhead
    \multicolumn{10}{c}{\ldots}
    \endfoot
    \endlastfoot
    \small \makecell {1} &
	\small \makecell {8h11m17s} &
	\small \makecell {$+25\degree12'29''$} &
	\small \makecell {0.85$\pm$0.82} &
	\small \makecell {0.95$\pm$0.20} &
	\small \makecell {0.39$\pm$0.20} &
	\small \makecell {2.53$\pm$0.30} &
	\small \makecell {0.37$\pm$0.20} &
	\small \makecell {0.66$\pm$0.22} &
	\small \makecell {0.24$\pm$0.21} \\
	\small \makecell {2} &
	\small \makecell {8h11m17s} &
	\small \makecell {$+25\degree12'06''$} &
	\small \makecell {2.62$\pm$0.77} &
	\small \makecell {1.16$\pm$0.25} &
	\small \makecell {2.82$\pm$0.26} &
	\small \makecell {4.48$\pm$0.21} &
	\small \makecell {0.26$\pm$0.20} &
	\small \makecell {0.93$\pm$0.22} &
	\small \makecell {0.31$\pm$0.22} \\
	\small \makecell {3} &
	\small \makecell {8h11m17s} &
	\small \makecell {$+25\degree10'35''$} &
	\small \makecell {4.06$\pm$0.91} &
	\small \makecell {2.11$\pm$0.25} &
	\small \makecell {3.65$\pm$0.26} &
	\small \makecell {7.27$\pm$0.32} &
	\small \makecell {0.50$\pm$0.30} &
	\small \makecell {0.73$\pm$0.33} &
	\small \makecell {0.54$\pm$0.33} \\
	\small \makecell {4} &
	\small \makecell {8h11m16s} &
	\small \makecell {$+25\degree10'40''$} &
	\small \makecell {0.82$\pm$0.50} &
	\small \makecell {0.67$\pm$0.16} &
	\small \makecell {0.67$\pm$0.16} &
	\small \makecell {2.09$\pm$0.16} &
	\small \makecell {0.05$\pm$0.16} &
	\small \makecell {0.35$\pm$0.17} &
	\small \makecell {0.43$\pm$0.17} \\
	\small \makecell {5} &
	\small \makecell {8h11m16s} &
	\small \makecell {$+25\degree10'42''$} &
	\small \makecell {0.93$\pm$0.56} &
	\small \makecell {0.39$\pm$0.16} &
	\small \makecell {0.63$\pm$0.16} &
	\small \makecell {2.01$\pm$0.35} &
	\small \makecell {0.27$\pm$0.23} &
	\small \makecell {0.46$\pm$0.25} &
	\small \makecell {0.46$\pm$0.25} \\
\hline
\end{longtable}
\clearpage
%\twocolumn

\begin{figure*}
    \includegraphics[scale=0.8]{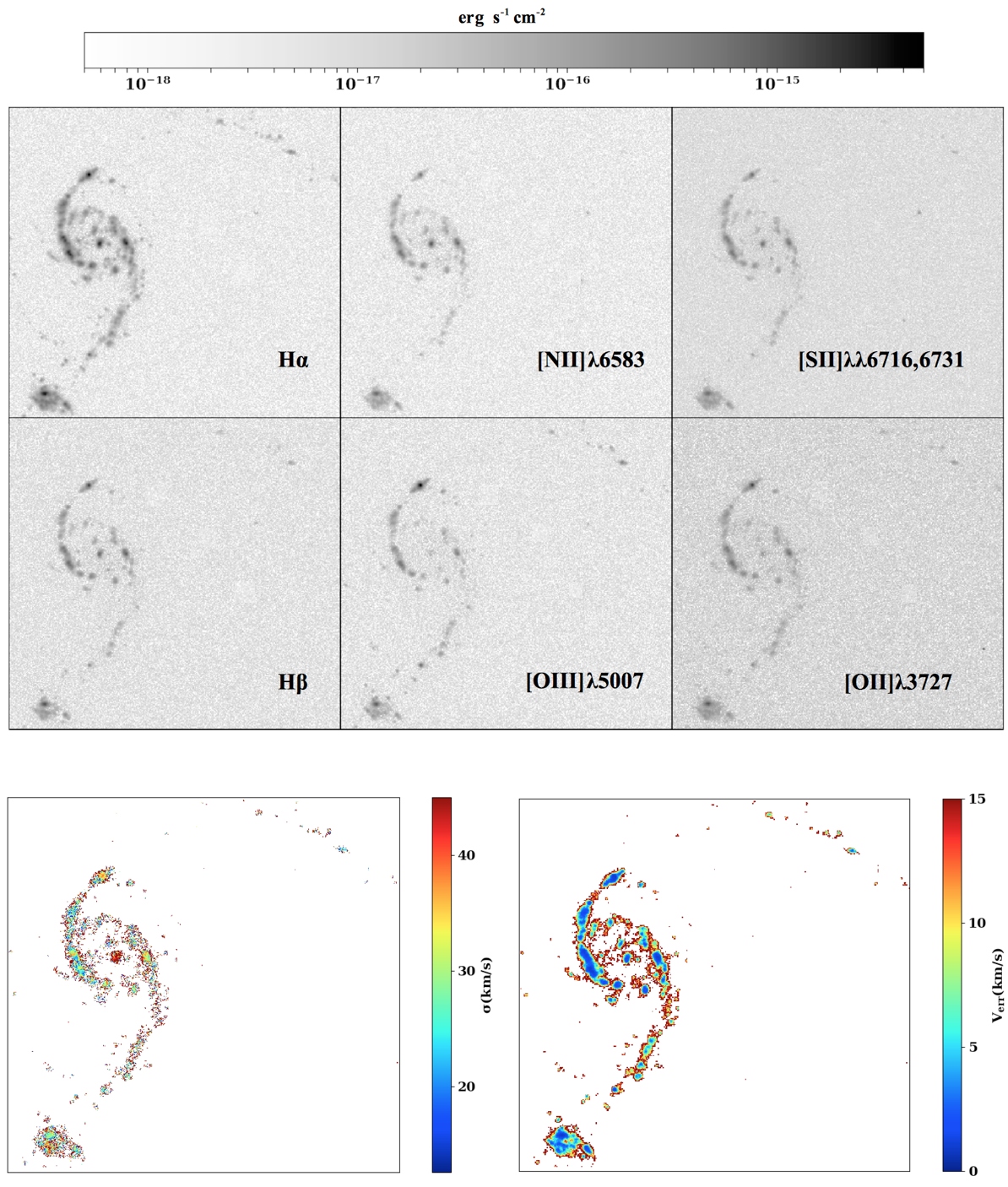}
    \caption{Top: Flux maps of spectral lines extracted from data cubes. Bottom: Velocity dispersion and velocity uncertainty maps ; only pixels with ${\rm H\alpha\ fluxes}>2\times10^{-17}\rm erg\,s^{-1}\,cm^{-2}$, $\rm SNR > 3$ and measured velocity uncertainties < 20 km s$^{-1}$ are displayed. }
    \label{fig:emissionlines_figure}
\end{figure*}

%%%%%%%%%%%%%%%%%%%%%%%%%%%%%%%%%%%%%%%%%%%%%%%%%%

% Don't change these lines
\bsp	% typesetting comment
\label{lastpage}
\end{document}